\documentclass[12pt,draftclsnofoot,onecolumn]{IEEEtran}
\usepackage{cite}

\usepackage{graphicx,color,epsfig,rotating,subfigure}
\usepackage{amsfonts,amsmath,amssymb}
\usepackage{algorithm}
\usepackage{subfigure}
\usepackage{algpseudocode}

\long\def\comment#1{}

\newfont{\bbb}{msbm10 scaled 700}

\newfont{\bb}{msbm10 scaled 1100}

\newcommand{\ZZ}{\mbox{\bb Z}}

\newcommand{\bv}{{\bf b}}

\newcommand{\hv}{{\bf h}}

\newcommand{\wv}{{\bf w}}

\newcommand{\xv}{{\bf x}}
\newcommand{\yv}{{\bf y}}
\newcommand{\zv}{{\bf z}}

\newcommand{\Am}{{\bf A}}
\newcommand{\Bm}{{\bf B}}

\newcommand{\Dm}{{\bf D}}

\newcommand{\Gm}{{\bf G}}
\newcommand{\Hm}{{\bf H}}
\newcommand{\Id}{{\bf I}}

\newcommand{\Qm}{{\bf Q}}

\newcommand{\Sm}{{\bf S}}

\newcommand{\Xm}{{\bf X}}
\newcommand{\Ym}{{\bf Y}}
\newcommand{\Zm}{{\bf Z}}

\newcommand{\Bc}{{\cal B}}
\newcommand{\Cc}{{\cal C}}

\newcommand{\Gc}{{\cal G}}

\newcommand{\Ic}{{\cal I}}

\newcommand{\Lc}{{\cal L}}

\newcommand{\Nc}{{\cal N}}

\newcommand{\Rc}{{\cal R}}
\newcommand{\Sc}{{\cal S}}
\newcommand{\Tc}{{\cal T}}

\renewcommand{\det}{{\hbox{det}}}

\newcommand{\SNR}{{\sf SNR}}
\newcommand{\INR}{{\sf INR}}

\newcommand{\eqdef}{\stackrel{\Delta}{=}}

\newcommand{\herm}{{\sf H}}

\newtheorem{definition}{Definition}
\newtheorem{theorem}{Theorem}
\newtheorem{lemma}{Lemma}
\newtheorem{corollary}{Corollary}

\newtheorem{remark}{Remark}

\begin{document}

\title{A Novel Cooperative Strategy for\\ Wireless Multihop Backhaul Networks}
\author{\authorblockN{Song-Nam Hong, Ivana Mari\'c and Dennis Hui}\\
\authorblockA{Ericsson Research, San Jose, CA, \\{\it email:(songnam.hong, ivana.maric, dennis.hui)@ericsson.com}}
}
\maketitle

\begin{abstract}
The 5G wireless network architecture will bring dense deployments of base stations called {\em small cells} for both outdoors and indoors traffic. The feasibility of their dense deployments depends on the existence of a high data-rate transport network that can provide high-data backhaul from an aggregation node  where data traffic originates and terminates, to every such small cell.  Due to the limited range of radio signals in the high frequency bands, multihop wireless connection may need to be established between each access node and an aggregation node. In this paper, we present a novel transmission scheme for wireless multihop backhaul for 5G networks. The scheme consists of 1) {\em group successive relaying} that established a relay schedule to efficiently exploit half-duplex relays and 2) an optimized quantize-map-and-forward (QMF) coding scheme that improves the performance of QMF and reduces the decoding complexity and the delay. We derive an achievable rate region of the proposed scheme and attain a closed-form expression in the asymptotic case for several network models of interests. It is shown that the proposed scheme provides a significant gain over multihop routing (based on decode-and-forward), which is a solution currently proposed for wireless multihop backhaul network. Furthermore, the performance gap increases as a network becomes denser. For the proposed scheme, we then develop energy-efficient routing that determines {\em groups} of participating relays for every hop. To reflect the metric used in the routing algorithm, we refer to it as {\em interference-harnessing} routing. By turning interference into a useful signal,  each relay requires a lower transmission power to achieve a desired performance compared to other routing schemes. Finally, we present a low-complexity successive decoder, which makes it feasible to use the proposed scheme in practice.
\end{abstract}

\section{Introduction}\label{sec:Intro}

To cope with the exponential increase in mobile data traffic, cellular networks are undergoing a paradigm shift from a well-planned deployment of tower-mounted base stations (BSs) to a capacity-driven (possibly ad hoc) deployment of smaller and lower-power BSs called {\em small cells}. This new architecture will cover both the outdoors and indoors scenarios and typically operate in higher frequency bands (mmW) where  more spectrum is available \cite{Pi}. While conventional BSs are typically connected to a high capacity point-to-point backhaul network, the same will not be possible for small cells due to their dense deployment at often more adverse locations. For such networks, the backhaul is a major bottleneck that can limit the throughput and deployment cost \cite{SmallCell}. To overcome this problem, backhaul  in 5G needs to adapt to the small cell architecture and bring new, more efficient solutions. Towards that goal, 5G backhaul is expected to be more integrated with the access network with respect to both technology and spectrum. In particular, wireless self-backhaul \cite{Hui,Interdigital,Baldemair} where the same radio spectrum is used for both access and transport, is an attractive solution that can substantially reduce the deployment cost, requiring only a single radio at each node and allowing for more flexibility in radio resource allocation.

\begin{figure}[t]
\centerline{\includegraphics[width=7cm]{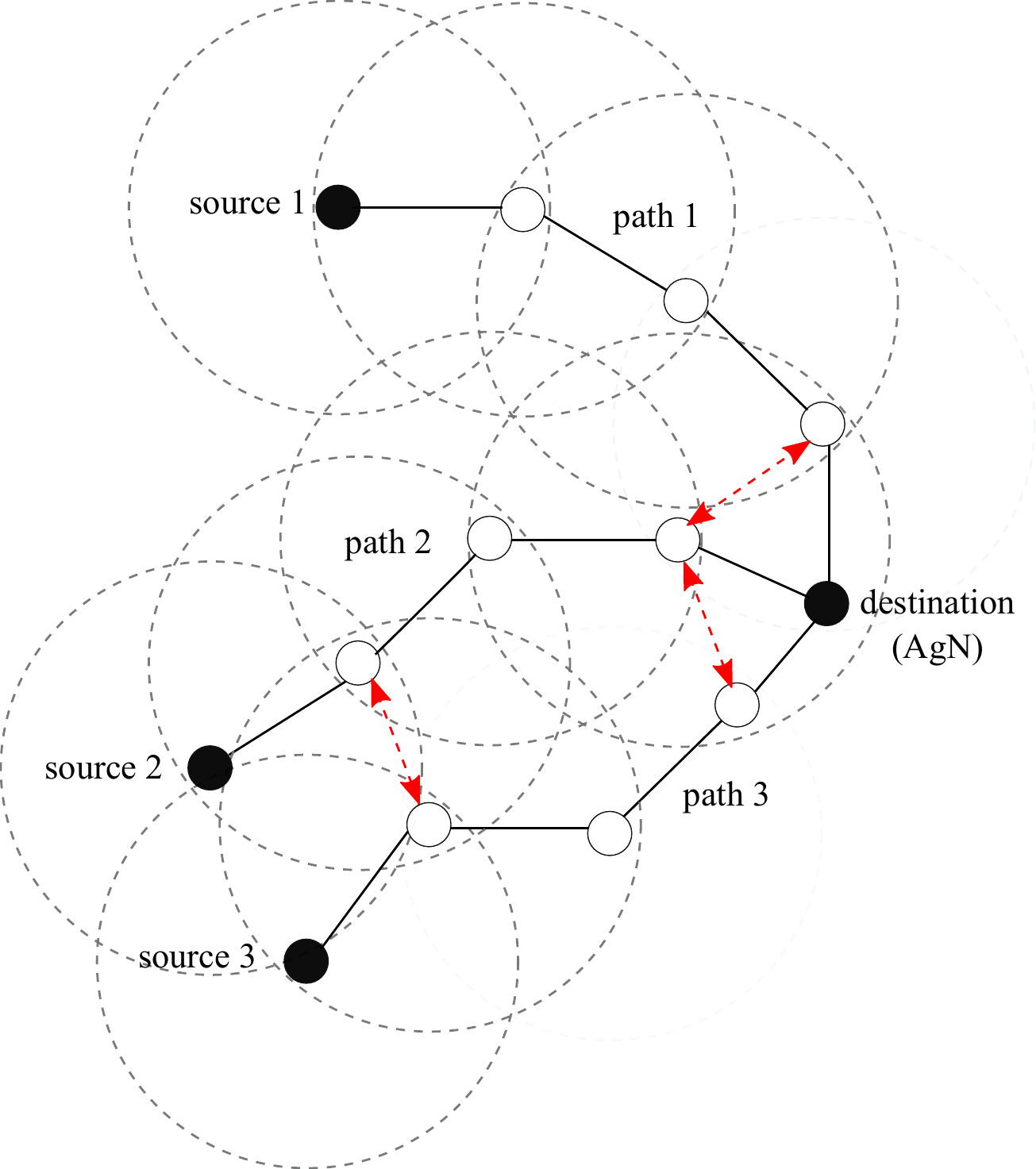}}
\caption{Routes obtained via a routing algorithm in an network with multiple sources and a single destination. A dashed circle represents a communication range whose radius can be determined as function of a transmit power. Red-dashed lines indicate strong interference. Any signal outside the communication range is relatively small and captured as additive Gaussian noise.}
\label{MR}
\end{figure}

Another key departure from the traditional backhaul is that, due to the limited range of radio signals at high frequencies \cite{Daniels}, a {\it multihop} wireless connection may need to be established between each access node (AN) and an aggregation node (AgN)\footnote{Notice that our work  can be immediately applied to the uplink of radio access networks (C-RANs) with a multihop backhaul network in \cite{Park} where sources, relays, and destination in Fig.~\ref{MR} correspond to mobile stations, radio units, and control units, respectively.} where an optical wired or high speed radio link connection is available (see Fig.~\ref{MR}). In wireless multihop backhaul, an access node serves not only its own assigned user terminals (UTs) in the vicinity, but also neighboring access nodes as a relaying node, in order to forward their data towards and/or from an AgN. Therefore, messages are routed (relayed) over multiple wireless hops to reach their destination.

\begin{figure*}[t]
\centerline{\includegraphics[width=16cm]{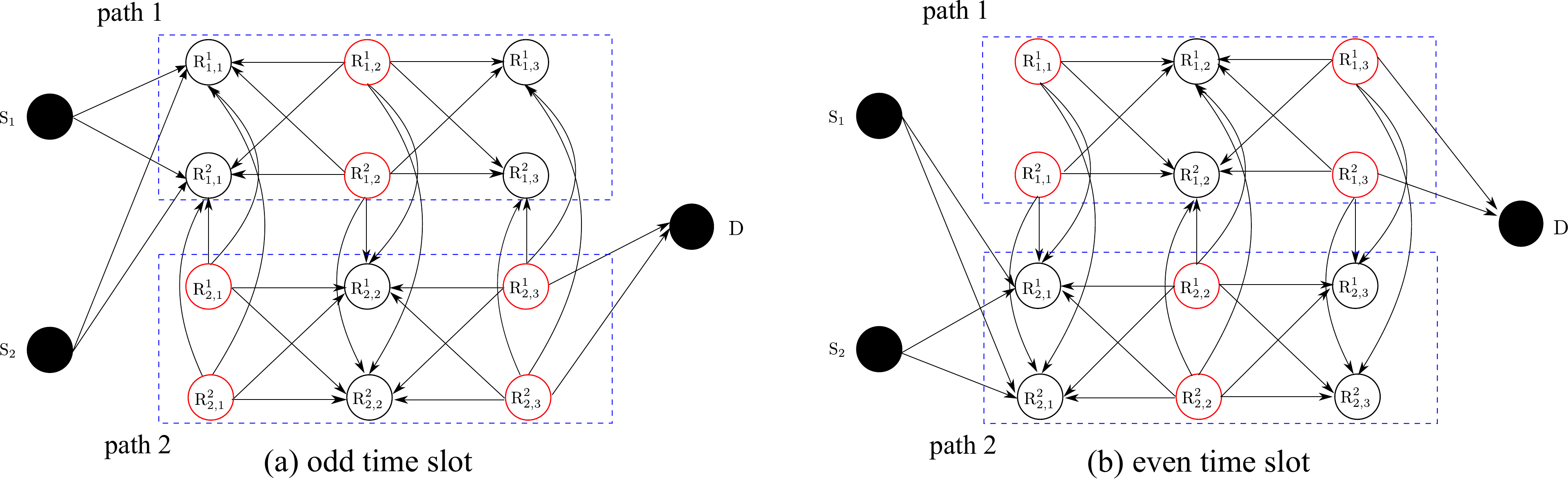}}
\caption{Group successive relaying for wireless backhaul networks when $L=2$ and $K=3$. In each time slot, relays that are transmitting are shown in red circles and relays that are receiving are shown in black circles. The sources transmit and the destination receives in every time slot. Each path contains the $L$ routes for $L$ sources.}
\label{m_layer}
\end{figure*}

Due to the broadcast nature of wireless medium, interference becomes a main limiting factor on the network throughput for wireless multihop backhaul network.
In \cite{Jain,Draves,Parissidis}, interference-aware routing was proposed and showed to offer a significant throughput gain over shortest-path routing.
A joint routing and resource allocation  for wireless backhaul networks was presented in \cite{Hui}.  Because  it is assumed that each relay decodes its desired message by treating other signals as noise, interference-aware routing aims at avoiding inter-route interference. It was shown in \cite{Hui} that this approach incurs significant limitation on network throughput at high load (i.e., the number of sources is large). This result is expected since it is nearly impossible to avoid all inter-route interference at high load. Furthermore, because the transmission rate on every route is determined by the minimum of all link-capacities on the route, strong interference on any one link of the route can drastically degrade the end-to-end performance (see Fig. 1). Instead, in this case, a more advanced coding scheme that can efficiently manage strong interference instead of simply treating it as noise should be considered. We develop such a scheme in this work.

Our proposed approach is motivated by the view of the wireless backhaul network as a multiple-multicast relay network. In such networks, the communication scheme that achieves the best known performance is quantize-map-and-forward (QMF) \cite{Avestimehr} (and a more general scheme of noisy network coding (NNC) \cite{Lim,Hou}). QMF/NNC achieves the cut-set upper bound (on the total network throughput) within a constant gap where the gap grows {\em linearly} with the number of relay nodes. This gap, however, may not be negligible for the system with multihop transmissions such as wireless backhaul network operating at high frequencies. In this paper, unlike \cite{Avestimehr,Lim,Hou}, we assume that each node operates in half-duplex mode, as is the case in many practical systems \cite{Peters,Peters1,Parkvall}, due to the problem of saturating the radio front-end when transmitting and receiving in the same frequency band and during the same time slot. This makes the considered problem more complicated since we need to determine the transmit/receive mode of each node for a given time slot. In fact, finding an optimal transmission strategy for wireless multihop backhaul networks with either full-duplex or half-duplex nodes is still an open problem.

In this paper, we present a novel transmission scheme for a wireless backhaul network where $L$ sources transmit their independent messages to the destination via $(K+1)$ hops.  For each hop, we form a relay stage that consists of $2L$ half-duplex relays. We refer to such network as a layered network. The layered network structure is formed via a routing algorithm  that we will optimize. In contrast to routing,   in our scheme, data from every source message is simultaneously forwarded by $L$ relays at each hop.  This is achieved by  {\em group successive relaying} and a coding scheme  that we develop and that uses the principle of QMF/NNC that we develop. In group successive relaying, $2L$ relays at each stage are divided into two groups of equal size so that while the relays in one group transmit their messages to the next stage, the relays in the other group receive signals from the previous stage. The role of each group is swapped at each time interval (see Fig.~\ref{m_layer}). In this way, the sources send $L$ new messages to the destination at every time slot. Hence, this relaying scheduling is optimal in the sense of maximizing the number of  data streams (i.e., sources' messages). Under the group successive relaying, we present an ``optimized" QMF which ensures a high rate per each message. In the conventional QMF \cite{Avestimehr}, each relay first quantizes its received signal at the background noise level, then randomly maps it to a Gaussian codeword, and transmits the codeword. The random mapping can be constructed by random binning and encoding the bin index with a Gaussian codeword \cite{Peyman}. While DF relays forward one of source messages to the next stage, QMF relays forward a function of all source messages (bin indices) as in network coding to the next stage. The destination does not explicitly decode the bin indices forwarded by relays. Instead, the source messages are decoded jointly with the bin indices, using the {\em time-expanded} network in Fig.~\ref{TimeEx}. While NNC scheme does not require time-expanded graph due to a vector quantization, it also performs joint decoding over multiple transmission blocks. This requires a joint decoder over the entire network in Fig.~\ref{TimeEx}, requiring a high computational complexity at destination. The proposed coding scheme (called ``Optimized" QMF) is different from the conventional QMF \cite{Avestimehr} in the four-fold: 1) Destination performs {\em forward} decoding in which it can start decoding a source message after one time slot, using a subnetwork (denoted by $\Sc_{i}$ in Fig.~\ref{TimeEx}) and the joint decoding is separately performed for each stage within a subnetwork, whereas it is done at once over the entire (time-expanded) network in \cite{Avestimehr,Lim}. Thus, the proposed QMF significantly reduces the decoding complexity and the delay; 2) Destination explicitly decodes all relays' messages (bin indices) and exploits them as side-information in the next time slot, which can completely eliminate interference among subnetworks; 3) A quantization level at each relay is optimized instead of choosing a background noise level; 4) Vector quantization is used as in \cite{Lim,Hui} whereas QMF uses scalar quantization. We show that using optimal quantization outperforms the other quantization methods that use background noise-level \cite{Avestimehr}, stage-depth \cite{Kolte},\footnote{In stage-depth quantization, the quantization level is chosen at a resolution decreasing with the number of stages $K$.} or classical Wyner-Ziv quantization \cite{Cover}. Furthermore, both using the stage-depth and optimal quantization attain better rate-scaling that the other two methods, i.e., the performance degrades {\em logarithmically} with $K$, thereby attaining a larger performance gain as $K$ grows. This result shows that optimizing quantization levels at relays substantially improves the rate-scaling and is indeed required for multihop transmissions.  Furthermore, we show that the proposed scheme provides a much higher rate than interference-aware routing based on decode-and-forward (in short, MR). The performance gap again increases as the network becomes denser (or load becomes higher). We further show that the proposed scheme has a {\em higher energy efficiency} compared to MR. The improved energy efficiency comes from enabling each relay to collect the signals broadcasted by transmitters (instead of treating them as noise). Based on these results, we propose  {\em interference-harnessing} routing in which the routing criterion is in contrast to the routing criterion of interference-aware routing \cite{Jain,Draves,Parissidis}: the routing metric is chosen to ``harness" interference, instead to avoid it.  Finally, to overcome the drawback of NNC which depends on prohibitively complex joint decoding at the destination, we develop a low-complexity stage-by-stage successive MIMO decoding, which can be implemented by using a conventional MIMO decoding (e.g., zero-forcing, linear MMSE, and integer-forcing receivers) instead of using complicated joint decoding.

The paper outline is as follows. In Section~\ref{sec:scheme}, we describe the proposed scheme and derive its achievable rate region. In Section~\ref{sec:Asymptotic}, asymptotic analysis is presented and a closed-form rate expression is derived. In Section~\ref{sec:EH}, we develop an interference-harnessing routing suitable for the proposed scheme. The low-complexity successive decoder is presented in Section \ref{sec:decoding}. In Section~\ref{sec:SIM}, numerical results for small networks are provided. Section~\ref{sec:conclusion} concludes the paper.

\section{The Proposed Transmission Scheme}\label{sec:scheme}

We consider a wireless backhaul network where $L$ sources want to send their messages to destination with the aid of intermediate half-duplex relays. The proposed scheme consists of three parts: i) a routing algorithm performed to establish a $(K+1)$-hop $2L$-layer network; ii) a relay scheduling referred to as {\em group successive relaying} applied to assign the transmit/receive mode of each half-duplex relay for a given time slot, which is optimal in the sense of maximizing the number of messages (simultaneously transmitted to destination); iii) the optimized QMF to ensure a high data rate per each message.

\subsection{Establishing a layered network}\label{subsec:layered}

A routing algorithm (e.g., interference-harnessing routing proposed in Section~\ref{sec:EH}) is performed to establish $L$ routes for $L$ sources. As shown in Fig.~\ref{MR}, the number of hops (or relay stages $K$) of each route is chosen according to the communication range and source-destination distance where the communication range can be determined as a function of the transmit power. For the convenience of explanation, it is assumed that each route is composed of $K$ relay stages (i.e., $(K+1)$ hops). However, the proposed scheme can be naturally extended into the so-called {\em asymmetric} layered network where the routes from different sources can have different number of hops due to the various source-destination distances. This case will be explained in Section~\ref{sec:EH}. To overcome the half-duplex constraint, each stage of a route is formed by two half-duplex relays, used in transmit and receive modes, such that while one relay transmits its signal to the next stage, the other relay receives a signal from the previous stage. Namely, the routing algorithm establishes two routes per each source. Therefore, a $(K+1)$-hop $2L$-layer network is produced as shown in Fig.~\ref{m_layer} for $L=2$ and $K=3$.

\subsection{Group Successive Relaying}\label{subsec:scheduling}

In group successive relaying, relays are divided into two groups depending on their transmit/receive mode such that
\begin{align*}
\Gc_{1}=& \left\{\{\mbox{R}_{1,2k}^{j}\}_{k=1}^{K_{1}}, \{\mbox{R}_{2,2k-1}^{j}\}_{k=1}^{K_{2}}: j=1,\ldots,L\right\}\\
\Gc_{2}=& \left\{\{\mbox{R}_{1,2k-1}^{j}\}_{k=1}^{K_{2}}, \{\mbox{R}_{2,2k}^{j}\}_{k=1}^{K_{1}}: j=1,\ldots,L\right\},
\end{align*}
where $K_{1} = \lfloor (K-1)/2 \rfloor$ and $K_{2} = \lceil(K-1)/2\rceil$, and $\mbox{R}_{i,k}^{j}$ represents the relay at stage $k$ on the $i$-th route of source $j$. At odd (resp. even) time slot (consisting of $n$ channel uses), the relays in $\Gc_{1}$ (resp. $\Gc_{2}$) operate in transmit mode and the relays in $\Gc_{2}$ (resp. $\Gc_{1}$) operate in receive mode. In this way, each source $j$ transmits $\underline{\wv}^{j}_{t} \in \{1,\ldots,2^{n r_{j,i}}\}$ at rate $r_{j,i}$ to the destination at time slot $t$, where $i=1$ for odd time slot $t$ and $i=2$ for even time slot $t$, and the destination can decode $L$ new messages $(\underline{\wv}^{1}_{t-K},...,\underline{\wv}^{L}_{t-K})$. Here, we used two different rates $r_{j,1}$ and $r_{j,2}$ per each source since the odd-indexed and even-indexed messages are conveyed to the destination via two disjoint paths: $\mbox{{\em path 1:}}  \left(({\rm S}_{1},\ldots,{\rm S}_{L}), (\mbox{R}_{1,1}^{j},\ldots,\mbox{R}_{1,K}^{j}:j=1,\ldots,L),{\rm D}\right)$ and
$\mbox{{\em path 2:}}  \left(({\rm S}_{1},\ldots,{\rm S}_{L}),(\mbox{R}_{2,1}^{j},\ldots,\mbox{R}_{2,K}^{j}:j=1,\ldots,L),{\rm D}\right)$.
 Notice that each path consists of $L$ routes for $L$ sources. As shown in Fig.~\ref{m_layer}, it is assumed that each transmit signal is received only at the neighboring relays in the received mode, namely, relays at the same stage on the other path and relays at the previous and next stages on the same path. This assumption may be reasonable since the strengths of signals from the outside of neighboring relays are relatively small thereby being able to be captured by additive Gaussian noise (see Fig.~\ref{MR}).

Incorporating the group successive relaying into a channel, the discrete memoryless channel is described by the transaction probabilities as
\begin{align}
&\prod_{j=1}^{L}\prod_{k=1}^{\lceil K/2 \rceil} p\left(y^{j}_{i,2k-1}|x^{j}_{i,2k-2},x^{1}_{\bar{i},2k-1},\ldots,x^{L}_{\bar{i},2k-1},x^{1}_{i,2k},\ldots,x^{L}_{i,2k}\right)\cdot\nonumber\\
&\;\;\;\;\;\prod_{k=1}^{\lfloor K/2 \rfloor} p\left(y^{j}_{\bar{i},2k}|x^{j}_{\bar{i},2k-1},x^{1}_{i,2k},\ldots,x^{L}_{i,2k},x^{1}_{\bar{i},2k+1},\ldots,x^{L}_{\bar{i},2k+1}\right)p\left(y_{{\rm D}}|x^{1}_{\bar{i},K},\ldots,x^{L}_{\bar{i},K}\right),\label{eq:echannel}
\end{align}where $i=1$ for odd time slot and $i=2$ for even time slot, and where $x_{i,k}^{j}$ and $y_{i,k}^{j}$ denote respective input and output at relay $\mbox{R}_{i,k}^{j}$, and $x_{1,0}^{j}$ and $x_{2,0}^{j}$ denotes source $j$'s inputs.

\begin{figure}
\centerline{\includegraphics[width=14cm]{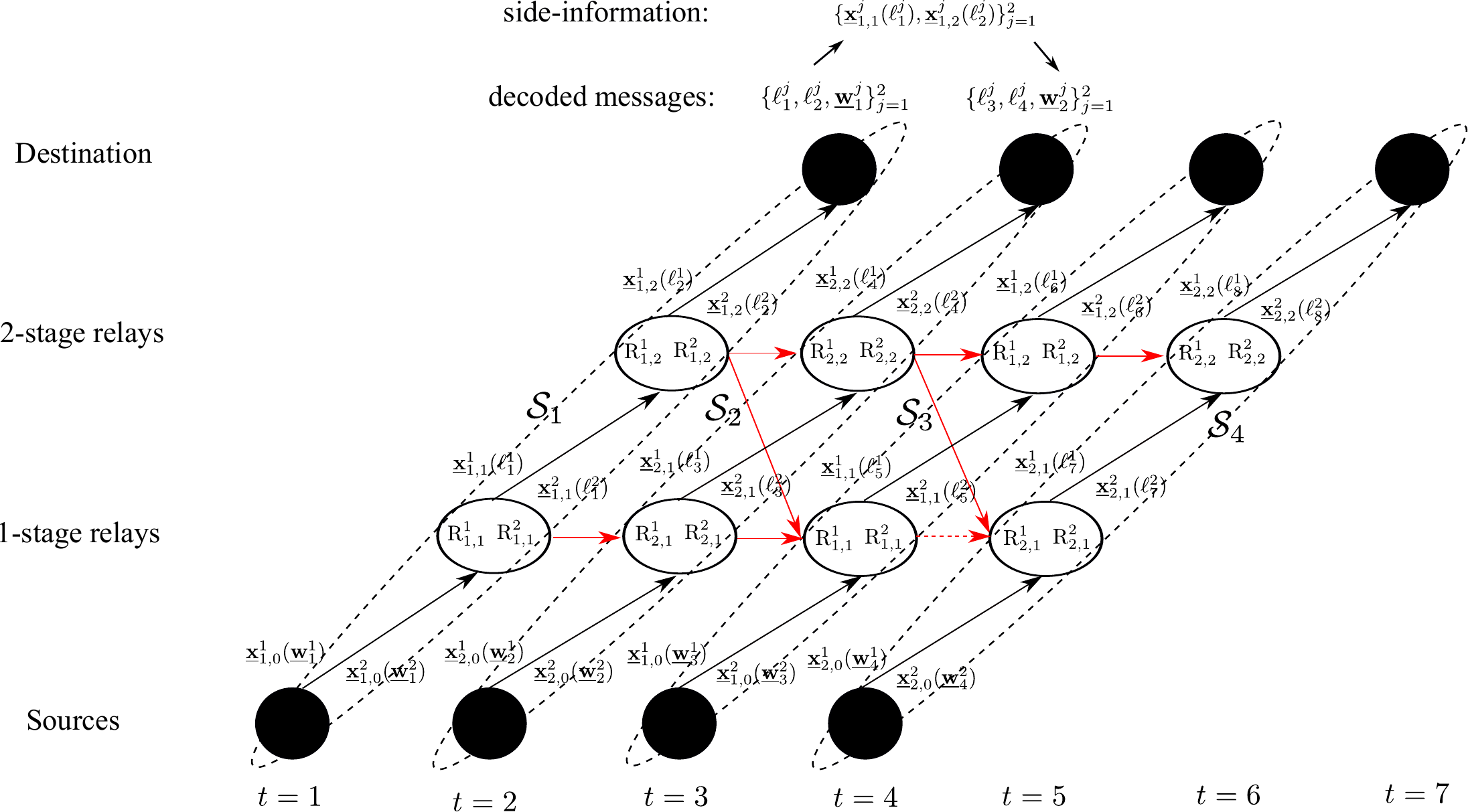}}
\caption{Time expanded multihop network with $L=2$ and $K=2$ where relays in transmit mode are shown at each time slot. Red lines stand for the ``known" interferences at destination and each arrow actually represents the wireless MIMO channel with two inputs and two outputs.}
\label{TimeEx}
\end{figure}

\subsection{Optimized QMF}\label{subsec:coding}

We present a novel coding scheme for wireless backhaul networks using group successive relaying as the underlying relay scheduling. Accordingly, the channel model in (\ref{eq:echannel}) is considered. Fig.~\ref{TimeEx} shows the {\em time-expanded} network for the case of $L=2$ and $K=2$. This will be used for the decoding of the proposed QMF scheme to deal with cycles in the original network as in \cite{Avestimehr}. To clarify our contribution on coding scheme, we compare the proposed scheme (named ``optimized" QMF) with the previous schemes of QMF, NNC, and SNNC:

i) QMF and NNC in \cite{Avestimehr,Lim} consist of message repetition encoding (i.e., one long message with repetitive encoding), signal quantization at relay, and simultaneous joint decoding on the received signals from all time slots (say, $t=1,2,\ldots,T$) without explicitly decoding relays' messages (i.e., quantization indices).

ii) SNNC in \cite{Hui} overcomes the long delay of NNC, by transmitting many short messages in time slots rather than using one long message with repetitive encoding. Instead of simultaneous joint decoding, destination performs {\em backward} decoding. After $T$ time slots, the destination starts decoding the source message $\underline{\wv}_{t}$ and relay's messages in the slot $t$, for $t=T,T-1,\ldots,1$, in descending order. Notice that joint decoding is separately performed for each slot, instead of jointly over all time slots.

iii) The optimized  QMF also uses short messages as in SNNC but performs {\em forward} decoding. The time-expanded network is partitioned into $T$ subnetworks, each of them is denoted by $\Sc_{t}$ for $t=1,\ldots,T$, as shown in Fig.~\ref{TimeEx}. After time slot $1$, the destination decodes the messages i.e., source message $\underline{\wv}_{1}$ and relays' messages) in the subnetwork $\Sc_{1}$, and then, after time slot $2$, decodes the messages in $\Sc_{2}$ by exploiting the previously decoded messages as side-information. Due to the use of successive relaying, the destination can completely know the interference seen at relays (depicted by red arrows in Fig.~\ref{TimeEx} since they are fully determined by the previously decoded relays' messages. In this way, the destination can decode the all source messages $\underline{\wv}_{t}$ for $t=1,\ldots,T$, in that order. Thus, the forward decoding has a lower decoding delay than the backward decoding in \cite{Hui}. Furthermore, it enables to find an optimal quantization level at each relay, which will be explained later in Section~\ref{sec:Asymptotic}.

\begin{figure}[t]
\centerline{\includegraphics[width=8cm]{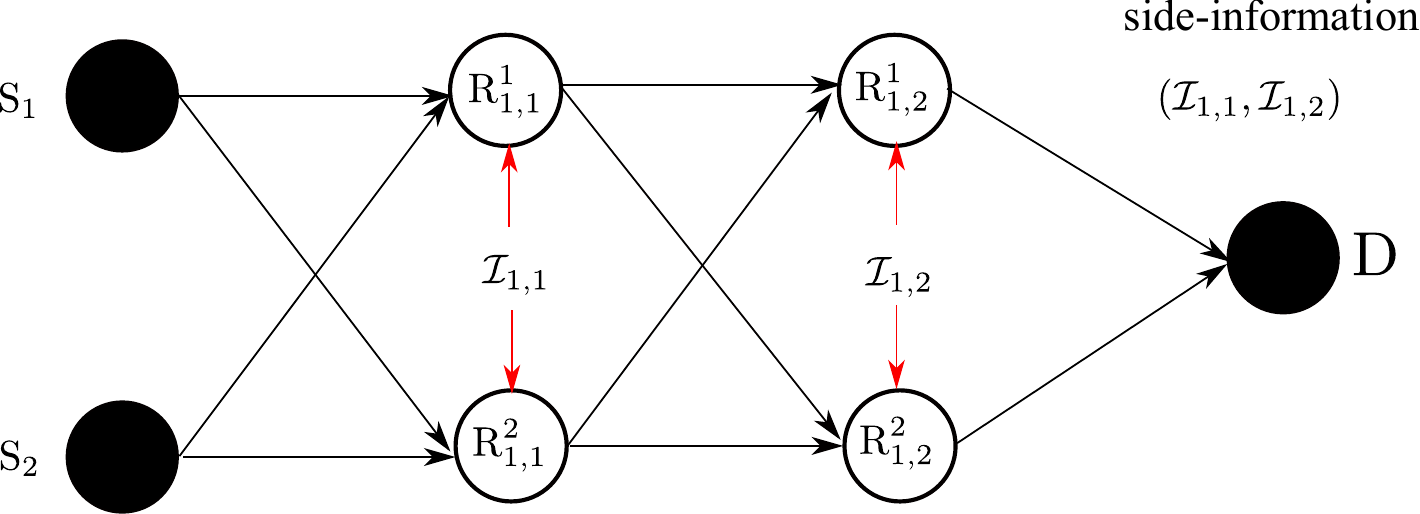}}
\caption{Equivalent simplified model of 3-hop 2-layer network where $\Ic_{1,K-1}\eqdef(\underline{\xv}_{2,K-1}^{1},\ldots,\underline{\xv}_{2,K-1}^{L},\underline{\xv'}_{1,K}^{1},\ldots,\underline{\xv'}_{1,K}^{L})$  denotes the known interference at destination.}
\label{m_simplified}
\end{figure}

Taking our decoding method (forward decoding) into account, we are able to produce the simplified channel model illustrated in Fig.~\ref{m_simplified} where we introduced the notation of $\underline{\xv'}_{i,k}^{j}$ since the known interference $\underline{\xv'}_{i,k}^{j}$ is different from $\underline{\xv}_{i,k}^{j}$ with respect to a time index. Since this model can be applied to decoding of any source message regardless of time indices, we will use this model to derive an achievable rate of the proposed scheme. For the ease of exposition, we introduce the notation $\Ic_{i,K-1}\eqdef(\underline{\xv}_{\bar{i},K-1}^{1},\ldots,\underline{\xv}_{\bar{i},K-1}^{L},\underline{\xv'}_{i,K}^{1},\ldots,\underline{\xv'}_{i,K}^{L})$ that denotes the known interference (i.e., side-information) at the destination. That is, the destination completely knows $(\Ic_{i,1},\ldots,\Ic_{i,K})$ when decoding source messages conveyed via path $i$.

From now on, we will explain the encoding/decoding procedures focusing on the case $i=1$ (i.e., for the network model in Fig.~\ref{m_simplified}). The same procedures apply to the case $i=2$. The encoding/decoding procedures are described as follows.

\textbf{Codebook generation:} Fix input distributions as
\begin{equation}
\prod_{i=1}^{2}\prod_{k=0}^{K}\prod_{j=1}^{L}p(x_{i,k}^{j})p(\hat{y}_{i,k}^{j}|y_{i,k}^{j}).\label{eq:inputdist}
\end{equation} Randomly and independently generate $2^{nr_{i}^{j}}$ codewords $\underline{\xv}_{i,0}^{j}(w_{i}^{j})$ of length $n$ indexed by $w_{i}^{j} \in \{1,\ldots,2^{n r_{i}^{j}}\}$ with i.i.d. components $\sim  p(x_{i,0}^{j})$. Randomly and independently generate $2^{n r_{i,k}^{j}}$ codewords $\underline{\xv}_{i,k}^{j}(\ell_{i,k}^{j})$ of length $n$ indexed by $\ell_{i,k}^{j} \in \{1,\ldots,2^{n r_{i,k}^{j}}\}$ with i.i.d. components $\sim p(x_{i,k}^{j})$. Independently generate $2^{n \hat{r}_{i,k}^{j}}$ codewords $\hat{\underline{\yv}}_{i,k}^{j}(\nu)$ of length $n$, indexed by $\nu \in \{1,\ldots,2^{n \hat{r}_{i,k}^{j}}\}$, with i.i.d. components $\sim p(\hat{y}_{i,k}^{j})$. The quantization codewords are randomly and independently assigned with uniform probability to $2^{n r_{i,k}^{j}}$ bins. We denote the $\ell_{i,k}^{j}$-th bin by $\Bc(\ell_{i,k}^{j})$ with $\ell_{i,k}^{j} \in \{1,\ldots,2^{n r_{i,k}^{j}}\}$.

\textbf{Encoding:} At every odd time slot, each source $j$ transmits a new message $\underline{\wv}^{j}_{1} \in \{1,\ldots,2^{nr_{1}^{j}}\}$ by sending the codeword $\underline{\xv}_{1,0}(\underline{\wv}^{j}_{1})$. Each relay ${\rm R}^{j}_{1,k}$ observes the $\underline{\yv}_{1,k}^{j}$ in receive mode. The relay quantizes its observation into a quantization codeword
$\hat{\yv}_{1,k}^{j}$ and finds a bin-index $\ell_{1,k}^{j} \in \{1,\ldots,2^{n r_{1,k}^{j}}\}$ such that the corresponding bin contains the quantization codeword $\hat{\yv}_{1,k}^{j}$. In transmit mode, the relay transmits the bin-index by sending the codeword $\underline{\xv}_{1,k}^{j}(\ell_{1,k}^{j})$.

\textbf{Stage-by-stage successive decoding:} At every time slot, the destination proceeds to decode $(\ell_{1,k}^{1},\ldots,\ell_{1,k}^{L})$ (i.e., the messages of relays at stage $k$) for $k=K,\ldots,1$, in that order, and then decode source messages $(w_{1}^{1},\ldots,w_{1}^{L})$. Notice that the destination knows $(\Ic_{1,1},\ldots,\Ic_{1,K})$ as side-information since as explained before, they are completely determined by the decoded messages in the previous time slot.

From the received observation $\underline{\yv}_{{\rm D}}$, the destination can reliably decode $(\ell_{1,K}^{1},\ldots,\ell_{1,K}^{L})$ if
\begin{align*}
(r_{1,K}^{1},\ldots,r_{1,K}^{L}) \in \Rc\left(p\left(y_{{\rm D}}|x_{1,K}^{1},\ldots,x_{1,K}^{L}\right),(\infty,\ldots,\infty)\right),
\end{align*} where the above rate region is equivalent to the MIMO MAC capacity region\footnote{In the MARC, if the link-capacity between each relay to destination is infinity, then the destination can observe the received signals at the relays without any distortion. Thus, this model is equivalent to the MIMO MAC channel.} and where $\Rc(\cdot)$ is given in Definition~\ref{def:region} below. Using the decoded bin-indices
$(\ell_{1,K}^{1},\ldots,\ell_{1,K}^{L})$ and the side-information $\Ic_{1,K}$, the destination performs {\em joint typical decoding} to decode $(\ell_{1,K-1}^{1},\ldots,\ell_{1,K-1}^{L})$, namely, it finds unique message $(\ell_{1,K-1}^{1},\ldots,\ell_{1,K-1}^{L})$ such that
\begin{align*}
&\left(\underline{\xv}_{1,K-1}^{1}(\hat{\ell}_{1,K-1}^{1}),\ldots,\underline{\xv}_{1,K-1}^{L}(\hat{\ell}_{1,K-1}^{L}), \hat{\underline{\yv}}_{1,K}^{1}(\nu_{1,K}^{1}),\ldots,\hat{\underline{\yv}}_{1,K}^{L}(\nu_{1,K}^{L}),
\Ic_{1,k} \right)\in \Tc_{\epsilon}^{(n)}(X_{1,0}, \hat{Y}_{1,1}, \Ic_{1,K})\\
&\mbox{for some} \;\;  \hat{\underline{\yv}}_{1,K}^{j}(\nu_{1,K}^{j}) \in \Bc(\ell_{1,K}^{j}),
\end{align*} where $\Tc_{\epsilon}^{(n)}(X_{1,0}, \hat{Y}_{1,1}, \Ic_{1,K})$ denotes the jointly typical set for $p(x_{1,0},\hat{y}_{1,1},\Ic_{1,K})$.  For fixed $(r_{1,K}^{1},$ $\ldots,r_{1,K}^{L})$, the channel for decoding the above messages can be seen as the multiple access relay channel (MARC) consisting of $L$ sources, $L$ intermediate relays, and one destination. Specifically, $L$ sources transmit their signals to $L$ relays via wireless channel defined as
\begin{equation*}
\prod_{j=1}^{L} p\left(y_{1,K}^{j}|x_{1,K-1}^{1},\ldots,x_{1,K-1}^{L},\Ic_{1,K}\right),
\end{equation*} and  $L$ relays send their messages to the destination via noiseless links of capacities
$(r_{1,K}^{L},\ldots,r_{1,K}^{L})$. From Lemma~\ref{lem:MARC} and Definition~\ref{def:region} below, we have:
\begin{align*}
(r_{1,K-1}^{1},\ldots,r_{1,K-1}^{L})\in \Rc\left( \prod_{j=1}^{L} p\left(y_{1,K}^{j}|x_{1,K-1}^{1},\ldots,x_{1,K-1}^{L},\Ic_{1,K}\right),(r_{1,K}^{1},\ldots,r_{1,K}^{L})\right).
\end{align*}
With the same argument, the destination can decode $(\ell_{1,K-2}^{1},\ldots,\ell_{1,K-2}^{L})$ using the decoded bin-indices $(\hat{\ell}_{1,K-1}^{1},\ldots,\hat{\ell}_{1,K-1}^{L})$ and the side-information $\Ic_{1,K-1}$ if
\begin{align*}
(r_{1,K-2}^{1},\ldots,r_{1,K-2}^{L})\in \Rc\left(\prod_{j=1}^{L} p\left(y_{1,K-1}^{j}|x_{1,K-2}^{1},\ldots,x_{1,K-2}^{L},\Ic_{1,K-1}\right),(r_{1,K-1}^{1},\ldots,r_{1,K-1}^{L})\right).
\end{align*}
In general, the following rate-constraints should be satisfied to reliably decode all relays' messages: For $k=K,\ldots,1$,
\begin{align*}
(r_{1,k}^{1},\ldots,r_{1,k}^{L})\in \Rc\left(\prod_{j=1}^{L} p\left(y_{1,k+1}^{j}|x_{1,k+1}^{1},\ldots,x_{1,k+1}^{L},\Ic_{1,k+1}\right),(r_{1,k+1}^{1},\ldots,r_{1,k+1}^{L})\right).
\end{align*} Notice that the above rate regions can be derived using $K +1$ {\em recursively} defined channels. Finally, the destination can reliably decode $L$ new messages $(\underline{\wv}_{1}^{1},\ldots,\underline{\wv}_{1}^{L})$ using the decoded bin-indices $(\ell_{1,1}^{1},\ldots,\ell_{1,1}^{L})$ and the side-information $\Ic_{1,1}$ if
\begin{equation*}
(r_{1}^{1},\ldots,r_{1}^{L}) \in \Rc\left(\prod_{j=1}^{L} p\left(y_{1,1}^{j}|x_{1,0}^{1},\ldots,x_{1,0}^{L},\Ic_{1,1}\right),(r_{1,1}^{1},\ldots,r_{1,1}^{L})\right).
\end{equation*} The exactly same analysis can be applied to the case of $i=2$. From the above analysis and letting $r_{i}^{j} = r_{i,0}^{j}$ for the convenience of notation, we obtain:

\begin{theorem}\label{thm:rate} For the wireless backhaul network in Fig.~\ref{m_layer}, a rate tuple
$(r_{1}^{1},\ldots,r_{1}^{L},r_{2}^{1},\ldots,r_{2}^{L})$ is achievable if
\begin{align*}
(r_{i,k}^{1},\ldots,r_{i,k}^{L})\in \Rc\left(\prod_{j=1}^{L} p(y_{i,k+1}^{j}|x_{i,k}^{1},\ldots,x_{i,k}^{L},\Ic_{i,k+1}),(r_{i,k+1}^{1},\ldots,r_{i,k+1}^{L})\right),
\end{align*} for $k=K,K-1,...,0$, with initial value $r_{i,K+1}^{j}=\infty$, $j=1,\ldots,L$, for any input distributions defined in (\ref{eq:inputdist}).
\hfill\IEEEQED
\end{theorem}

\begin{lemma}\label{lem:MARC} Consider the MARC defined as
\begin{equation*}
\left(\prod_{j=1}^{L} p\left(y_{i,k+1}^{j}|x_{i,k}^{1},\ldots,x_{i,k}^{L},\Ic_{i,k+1}\right), \left(r_{i,k+1}^{1},\ldots,r_{i,k+1}^{L}\right)\right).
\end{equation*} When $\Ic_{i,k+1}$ is known to the receiver,  a rate tuple $(r_{i,k}^{1},\ldots,r_{i,k}^{L})$ is achievable if for all $\Lc \subseteq \{0,\ldots,L-1\}$,
\begin{align*}
&\sum_{\ell \in \Lc} r_{i,k}^{\ell}\leq \min_{\Sc \subseteq [1:L]} \sum_{j \in \Sc}\left[r_{i,k+1}^{j} - I\left(Y_{i,k+1}^{j};\hat{Y}_{i,k+1}^{j}|X_{i,k}^{1},\ldots,X_{i,k}^{L},\Ic_{i,k+1}\right)\right]\\
 &\;\;\;\;\;\;\;\;\;\;\;\;\;\;\;\;\;\;\;\;\;\;\;\;\;\;\;\;\;+ I\left((X_{i,k}^{\ell}:\ell \in \Lc);(\hat{Y}_{i,k+1}^{j}:j\in \Sc^{c})|(X_{i,k}^{j}:j\in \Lc^{c}),\Ic_{i,k+1}\right),
\end{align*} for any input distributions that factors into  $\prod_{j=1}^{L}p(x_{i,k}^{j})\prod_{j=1}^{L}p(\hat{y}_{i,k}^{j}|y_{i,k}^{j})$.
\end{lemma}
\begin{IEEEproof} The proof almost follows \cite[Appendix 1]{Sanderovich}. The considered model reduces to the model in \cite{Sanderovich} by setting $\Ic_{i,k+1}=\phi$. Since the known interference $\Ic_{i,k+1}$ can be completely canceled at the destination, it does not change the achievable rate region. This completes the proof.
\end{IEEEproof}

\begin{definition}\label{def:region} We denote the rate region of Lemma~\ref{lem:MARC} by
\begin{equation}
\Rc\left(\prod_{j=1}^{L} p\left(y_{i,k+1}^{j}|x_{i,k}^{1},\ldots,x_{i,k}^{L},\Ic_{k+1}\right),(r_{i,k+1}^{1},\ldots,r_{i,k+1}^{L})\right).
\end{equation}
\end{definition}

\begin{remark}\label{remark:CF} A classical compress-and-forward (CF)  \cite{Cover} can be considered as a special case of QMF in the sense that a distortion (or quantization) level at each relay is chosen according to Wyner-Ziv distortion \cite{Cover}. That is, the quantization level at relay $\mbox{R}_{i,k}^{j}$ is chosen such that
\begin{equation}\label{eq:Wyner}
I(\hat{Y}_{i,k}^{j};Y_{i,k}^{j}|\Ic_{i,k}) = r_{i,k}^{j}.
\end{equation} Then, the destination can avoid the complexity of joint typical decoding and just use classical successive decoding. Namely, the destination can find quantized observations using the decoded bin-indices and side-information. Then, it just performs ``classical" MIMO decoding based on quantized observations, which significantly reduces the decoding complexity at the destination. Thus, using Wyner-ziv quantization may be better choice in practice as long as it provides a satisfactory performance.
\end{remark}

\begin{remark} In Theorem~\ref{thm:rate}, we derived an achievable rate region of the proposed scheme. However, it is not so clear how to optimize relays' rates subject to their achievable rate region for maximizing a sum rate or a symmetric rate. This is non-trivial optimization problem and is interesting subject of a future work. In the next section, we partially solve this problem by restricting our attention to {\em symmetric} Gaussian networks.
\end{remark}

\section{Gaussian Wireless Multihop Backhaul Networks:\\ Asymptotic Analysis and Closed-Form Rate Expression}\label{sec:Asymptotic}

We consider a Gaussian wireless backhaul network where destination is equipped with $L$ receiver antennas and accordingly, the $L$-dimensional observations are denoted by $(\underline{\yv}_{i,K+1}^{1},\ldots,$ $\underline{\yv}_{i,K+1}^{L})$. Let $\Hm_{i,k}$ denote the $L \times L$ channel matrix with inputs $(\underline{\xv}_{i,k}^{1},\ldots,\underline{\xv}_{i,k}^{L})$ and outputs $(\underline{\yv}_{i,k+1}^{1},\ldots,\underline{\yv}_{i,k+1}^{L})$ for $k=0,1,\ldots, K$. Let $\Gm_{i,k}$ denote the $L \times L$ {\em intra-path interference} channel matrix with inputs $(\underline{\xv}_{i,k+1}^{1},\ldots,\underline{\xv}_{i,k+1}^{L})$ and outputs $(\underline{\yv}_{i,k}^{1},\ldots,\underline{\yv}_{i,k}^{L})$ for $k=1,\ldots,K-1$. Also, let $\Sm_{i,k}$ denote the $L \times L$ {\em inter-path interference} channel matrix with inputs $(\underline{\xv}_{\bar{i},k}^{1},\ldots,\underline{\xv}_{\bar{i},k}^{L})$ and outputs $(\underline{\yv}_{i,k}^{1},\ldots,\underline{\yv}_{i,k}^{L})$ for $k=1,\ldots,K$. From Theorem~\ref{thm:rate}, we can see that the performance of the proposed scheme is independent from the intra-path and inter-path interference matrices $\Gm_{i,k}$ and $\Sm_{i,k}$. For the Gaussian channel, Lemma~\ref{lem:MARC} is simplified as
\begin{lemma}\label{lem:G-MARC} For the Gaussian MARC defined by $\left(\Hm_{i,k},(r_{i,k+1}^{1},\ldots,r_{i,k+1}^{L})\right)$, a rate tuple $(r_{i,k}^{1},\ldots,$ $r_{i,k}^{L})$ is achievable for all $\Lc \subseteq \{0,\ldots,L-1\}$,
\begin{align}
\sum_{\ell \in \Lc} r_{i,k}^{\ell} &\leq \min_{\Sc \subseteq [1:L]} \sum_{j \in \Sc}\left[r_{i,k+1}^{j} - \log\left(1+1/Q_{i,k}^{j}\right)\right]+\log\det\left(\Id+\Dm_{i,k}\Hm_{i,k}(\Sc^{c})\Hm_{i,k}(\Sc^{c})^{\herm}\right),
\end{align} for some quantization level $Q_{i,k}^{j} \geq 0$, where $\Hm_{i,k}(\Sc)$ represents the channel sub-matrix to contain the rows of $\Hm_{i,k}$ with their indices belong to $\Sc \subseteq [1:L]$, $\Dm_{i,k}$ represents a diagonal matrix whose $j$-th diagonal element is $P_{{\rm tx}}/(1+Q_{i,k}^{j})$, and $P_{{\rm tx}}$ denotes a transmit power at each node.\hfill\IEEEQED
\end{lemma}

We evaluate the performance of the proposed scheme for two extreme scenarios of network structure called {\em sparse} and {\em dense} networks. They are classified according to the structure of $\Hm_{i,k}$. In a sparse network, the number of interfering nodes is constant as the number of nodes $L$ grows (i.e., $\Hm_{i,k}$ forms a band matrix) while in dense network, it increases linearly with $L$ (i.e., $\Hm_{i,k}$ forms a ``full" matrix). Although we limit our attention to Gaussian networks, the computation of the achievable rate in Lemma~\ref{lem:G-MARC} is generally difficult since it involves a complicated combinatorial optimization. In order to make a problem manageable, we consider the particular structures of $\Hm_{i,k}$ that captures the features of sparse and dense networks, respectively. They are described in Section~\ref{subsec:SN} and~\ref{subsec:DN}, respectively. Further, we let $L \rightarrow \infty$ (i.e., asymptotic analysis) and focus on the achievable symmetric rate of $r$.

\subsection{Sparse Networks}\label{subsec:SN}

In this section we assume that each transmit signal is received at its desired relay and two neighboring relays, i.e., $\Hm_{i,k}$ has the form of tri-diagonal structure with $(\alpha,1,\alpha)$, where $\alpha$ denotes the relative interference strength. Namely, other interferences are relatively weak and are captured by additive Gaussian noises. This model is well-known as Wyner model \cite{Wyner}. Obviously, this model is the form of sparse network in that the number of interfering nodes is fixed by $2$, independently from the number of users $L$. The transmit power at each node is assumed to be  $P_{{\rm tx}}=\SNR$. Hereafter, we restrict ourselves to choose relays' rates as $r_{k}=r_{i,k}^{j}$ for all $i$ and $j$, which is reasonable due to the symmetric structure of $\Hm_{i,k}$.  Then, by symmetry and concavity, this limits the sum-rate inequality to be dominant in Lemma~\ref{lem:G-MARC} and by choosing the same quantization levels at each relay (i.e., $Q_{k}=Q_{i,k}^{j}$ for all $i$ and $j$), we have:
\begin{align}
r_{k-1} &= \frac{1}{L} \min_{\Sc \subset [1:L]} |\Sc|\left(r_{k} - \log\left(1+\frac{1}{Q_{k}}\right)\right)+ \log\det\left(\Id+\left(\frac{P_{{\rm tx}}}{1+Q_{k}}\right)\Hm_{i,k}(\Sc^{c})\Hm_{i,k}^{\herm}(\Sc^c)\right).\label{eq:const-G}
\end{align}  A remarkable result of \cite{Sanderovich} is that in the limit of $L \rightarrow \infty$, (\ref{eq:const-G}) can be simplified to (\ref{eq:sim1}) below. Consequently, we get the following:

\begin{corollary} The proposed scheme achieves the symmetric rate (per user) $r=r_{0}$ to satisfy:
\begin{align}
r_{k-1} = F(x^{*}), \label{eq:sim1}
\end{align}for $k=K+1,\ldots,1$, where
\begin{align*}
F(x) = \int_{0}^{1} \log\left(1+\SNR(1-2^{-x})(1+2\gamma\cos(2\pi\theta))^2)\right)d\theta,
\end{align*} and $x^{*}$ is the solution of the equation $F(x) = r_{k} - x$ with the initial value $r_{K+1}=\infty$. \flushright$\blacksquare$
\end{corollary}

For the comparison, we consider the MR where each relay decodes its desired message (forwarded by its own route) by treating all other signals as noise. In order to derive its achievable rate, we assume that $\Gm_{i,k}=\mbox{0}$ and $\Sm_{i,k}=\mbox{0}$. Thus, the achievable rate of the MR is optimistic since in practice, they may not be zero matrices. Under this assumption, the achievable symmetric rate of the MR is given by
\begin{equation}
r_{{\rm  MR}} = \log\left(1+\frac{\SNR}{1+2\alpha^2\SNR}\right).
\end{equation}

\begin{figure}[t]
\centerline{\includegraphics[width=10cm]{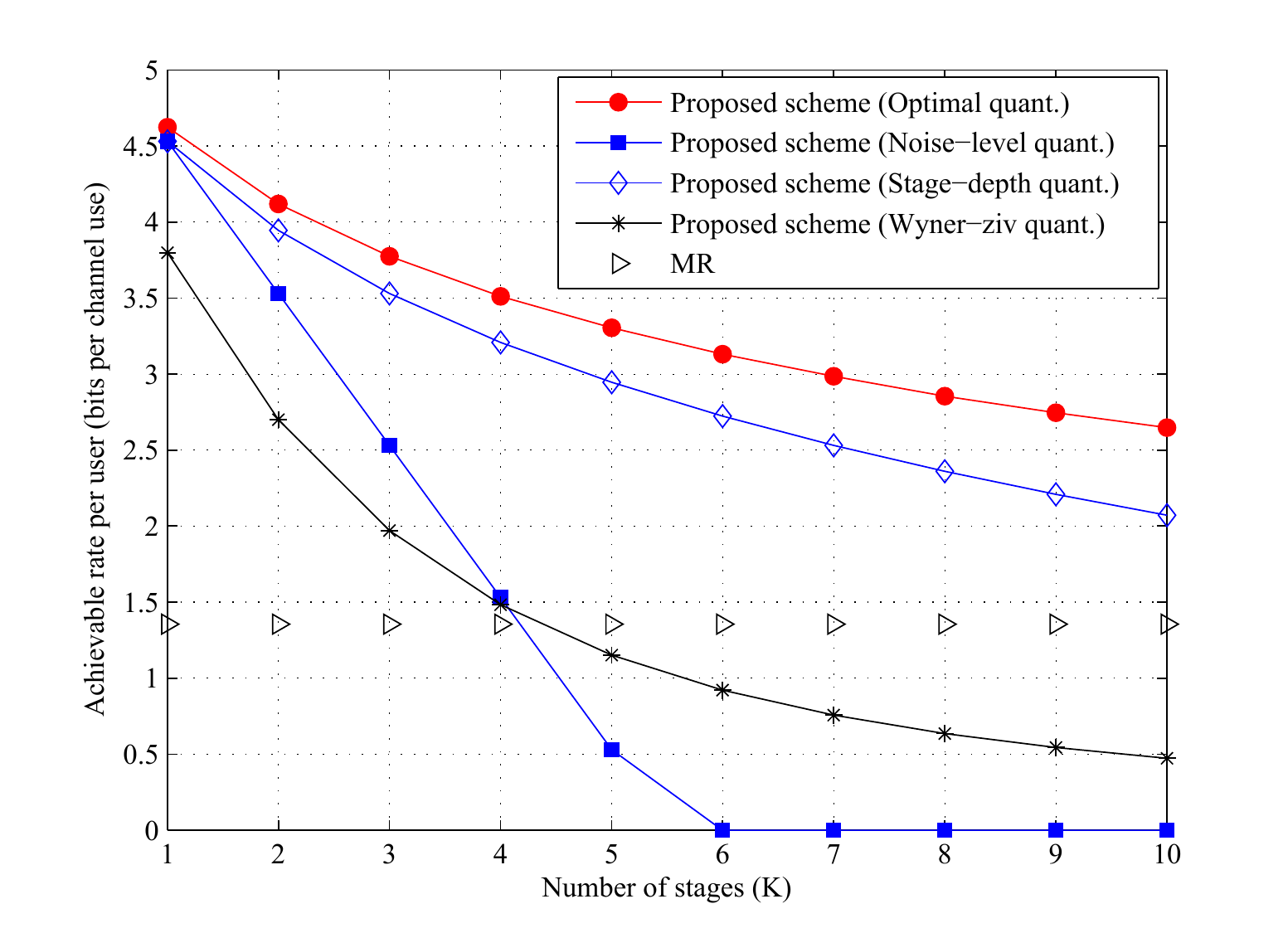}}
\caption{$\SNR=20$dB and $\INR=15$dB (e.g., $\alpha=\sqrt{\INR/\SNR}=0.56$). Performance of the proposed scheme for various quantization levels as a function of the number of stages $K$.}
\label{SIM-SN}
\end{figure}

Fig.~\ref{SIM-SN} shows the achievable symmetric rate of the proposed scheme for various quantization levels at relays. We first observe that using optimal quantization outperforms the other cases as  background noise-level (i.e., $Q_{k}=1$) \cite{Avestimehr}, stage-depth (i.e., $Q_{k}=K$) \cite{Kolte}, and Wyner-Ziv quantization (obtained from (\ref{eq:Wyner})) \cite{Cover}. As anticipated, the optimal and stage-depth quantization provide a larger performance gain due to the improved rate-scaling. When $K \geq 4$, the noise-level and Wyner-Ziv quantization show worse performance than the MR and hence optimizing a quantization level plays an important role for multihop networks.

\subsection{Dense Networks}\label{subsec:DN}

In this section we assume that $\Hm_{i,k}$ is an i.i.d. matrix whose entry has a complex Gaussian distribution with zero mean and unit variance. Note that the transmit power at each node is assumed to be $P_{{\rm tx}}=\SNR/L$. If we choose the $P_{{\rm tx}}=\SNR$ as in the sparse network, the received power at each relay goes to infinity, thereby yielding an infinite rate. As done in Section~\ref{subsec:SN}, we limit our attention to the symmetric choice of $r_{k}=r_{i,k}^{j}$ for all $j$ and $i$. With the same argument in Section~\ref{subsec:SN}, the sum-rate inequality in Lemma~\ref{lem:G-MARC} is dominant and is represented as in (\ref{eq:const-G}) with $P_{{\rm tx}}=\SNR/L$. A closed-form expression of (\ref{eq:const-G}) has been derived in \cite{Hong-SC} by using the fact that as $L \rightarrow \infty$, the problem becomes symmetric although the channel is ``full" and non tri-diagonal as in the Wyner model. The closed-form expression was obtained from the asymptotic random matrix theory  and the submodular structure of the rate expression, which equal to (\ref{eq:QMF}) below. Consequently, we get the following:

\begin{corollary}\label{cor:mLayer-G} The proposed scheme achieves the symmetric rate (per user) $r=r_{0}$ to satisfy:
\begin{eqnarray}
r_{k-1}=\min\left\{r_{k}-\log\left(1+\frac{1}{Q_{k}}\right),\Cc\left(\frac{\SNR}{1+Q_{k}}\right)\right\},\label{eq:QMF}
\end{eqnarray}for $k=K+1,\ldots,1$, with initial value $Q_{K+1}=0$ and $r_{K+1}=\infty$, where
\begin{equation}
\Cc(x) \eqdef 2\log\left(\frac{1+\sqrt{1+4x}}{2}\right)-\frac{\log{e}}{4x}(\sqrt{1+4x}-1)^2.
\end{equation} \flushright$\blacksquare$
\end{corollary}

Since the achievable rate in (\ref{eq:QMF}) is the minimum of two terms, where the first is an increasing function of $Q_{k}$ and the second
is a decreasing function of $Q_{k}$, the optimal value of $Q_{k}$ is attained by solving
\[r_{k}-\log\left(1+1/Q_{k}\right)= \Cc\left(\SNR/(1+Q_{k})\right).\]
Letting
\[f(Q_{k}) \eqdef r_{k}-\log\left(1+1/Q_{k}\right) - \Cc\left(\SNR/(1+Q_{k})\right),\]
we can find  $Q_{k,{\rm min}} = 1/ (2^{r_{k}} -1)$ and $Q_{k,{\rm max}}=(1+\SNR)/(2^{r_{k}}-1)$ such that
$f(Q_{k,{\rm min}}) \leq 0$  and  $f(Q_{k,{\rm max}}) \geq 0$. This is because $Q_{k,{\rm min}}$ makes the first term of the minimum in (\ref{eq:QMF}) zero and $Q_{k,{\rm max}}$ is the Wyner-Ziv quantization, which makes the second term to attain the minimum in (\ref{eq:QMF}). Using bisection method, we can quickly find an optimal quantization level $Q_{k,{\rm opt}}$. This value will be used to plot the performance of the optimized QMF.

\begin{figure}[t]
\centerline{\includegraphics[width=10cm]{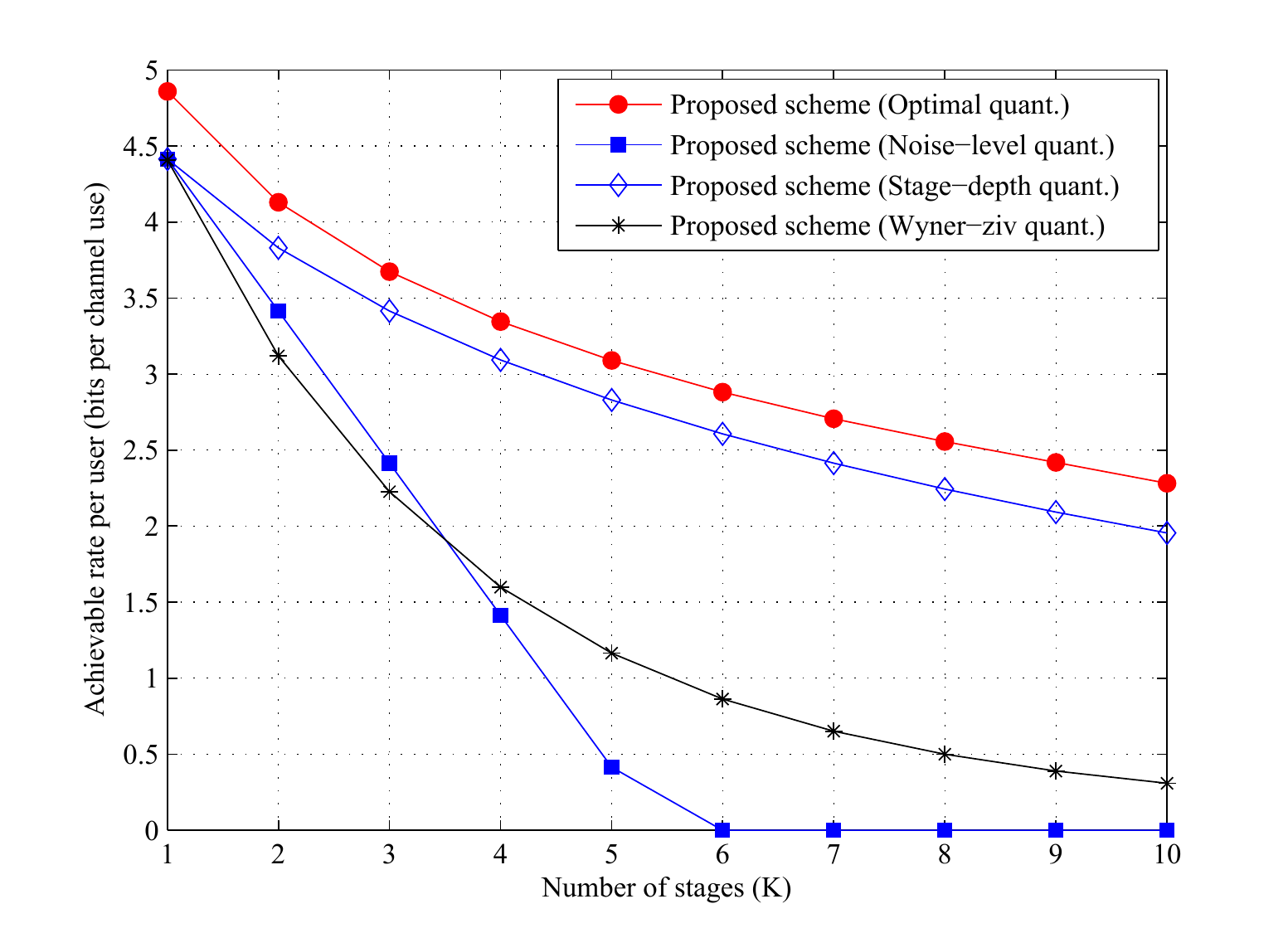}}
\caption{$L\SNR=20$ dB. Performance of the proposed scheme for various quantization levels as a function of the number of stages $K$. Notice that the performance of the MR goes to zero as $L \rightarrow \infty$ due to the impact of severe interference.}
\label{SIM-DN}
\end{figure}

Fig.~\ref{SIM-DN} shows the achievable symmetric rate of the proposed scheme for various quantization levels. The performance trend is similar as in the case of sparse network. The most remarkable observation is that the proposed scheme can achieve almost the same performance as in Fig.~\ref{SIM-SN} with a lower transmit power (per each node) as $L \rightarrow \infty$.

%
%
%
\section{Interference-Harnessing routing}\label{sec:EH}

The proposed scheme can be applied together with any multihop routing algorithm (e.g., interference-aware routing \cite{Hui}). In this section, we provide a routing criterion that further improves the performance of the proposed scheme. From Figs.~\ref{SIM-SN} and~\ref{SIM-DN}, we observe that the proposed scheme can achieve almost the same performance in both dense and sparse networks. In dense network, however, the transmit power consumption is reduced proportionally to the number of users $L$ (i.e., $P_{{\rm tx}}=\SNR/L$). In other words, the proposed scheme has a {\em higher energy efficiency} achieving a desired rate performance with a lower transmission power as the network becomes denser. The improved energy efficiency comes from enabling each relay (by using QMF) to collect the signals broadcasted by transmitters. Any interfering signal that is received at the relay will be forwarded through QMF and treated as a useful signal at the destination. For this reason, QMF performs better when the network is dense and interference is stronger. This motivates us to propose {\em interference-harnessing} routing in which the routing criterion contrasts the routing criterion used in the interference-aware routing. Interference-harnessing routing refers to a family of routing algorithms in which the metric is chosen to exploit interference. This can be done by, for example, choosing the metric that maximizes an achievable rate between every two consecutive relay stages. For any choice of relays at the two stages, this rate corresponds to a rate in a MIMO channel (see Section~\ref{sec:decoding}) and can thus be calculated, as we will specify in this section. An approximation of such metric would simply choose relays at each stage such that the signal power received from the previous stage at each of these relays is maximized (see Remark~\ref{remark:EH}). The performance gain of interference-harnessing  algorithm compared to the MR is demonstrated via simulation result in Section~\ref{sec:SIM}.

\begin{figure*}[t]
\centerline{\includegraphics[width=14cm]{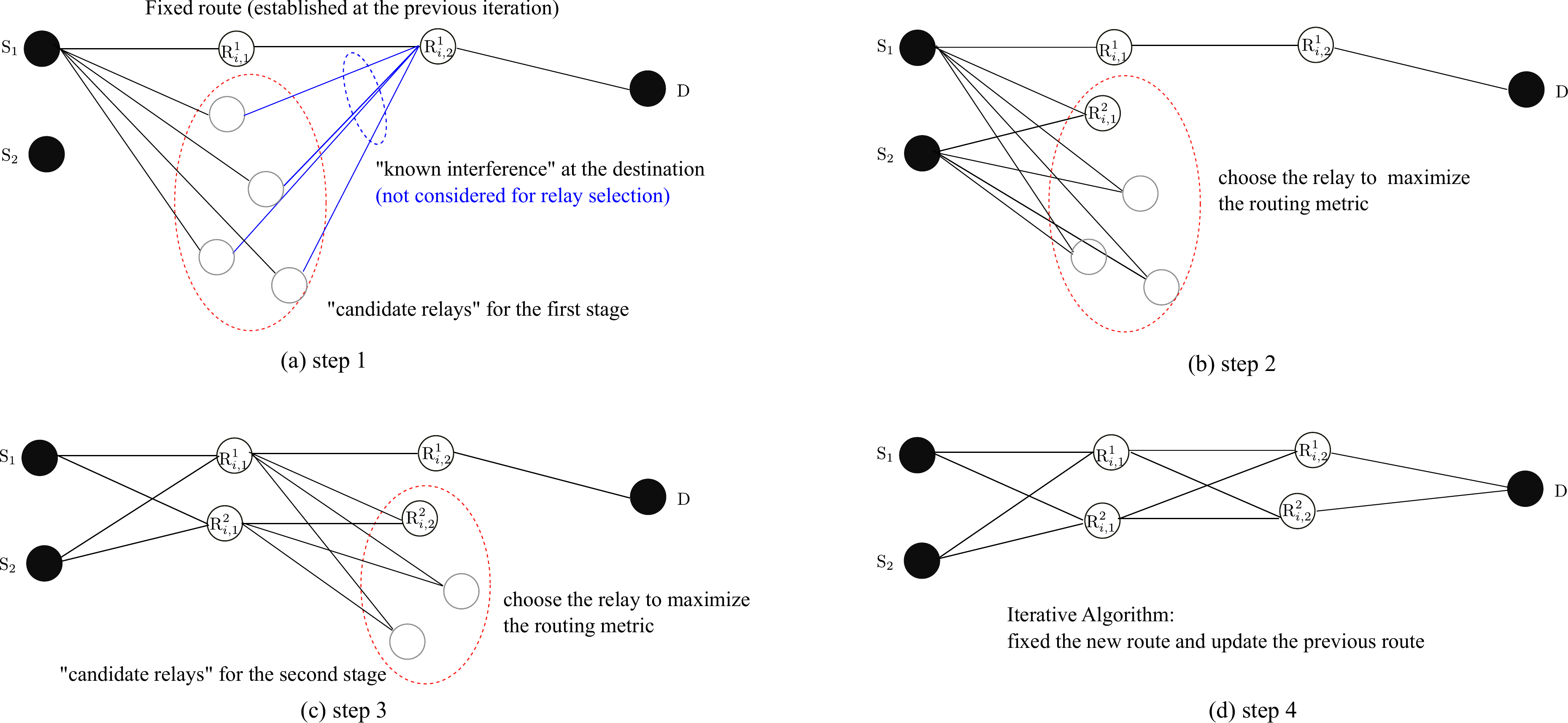}}
\caption{Illustration of an interference-harnessing routing to establish the two routes in path $i$ where ``candidate relays" denotes the relays within the communication range from a transmitter.}
\label{FIG-EH}
\end{figure*}

For the interference-harnessing routing, one efficient algorithm can be developed using the iterative algorithm in \cite{Hui} by properly modifying the routing criterion. In \cite{Hui}, the algorithm establishes one route at a time while keeping other previously established routes fixed, and repeats the process until negligible improvements in the sum throughput can be made. The routing criterion maximizes each link-capacity on the route which is computed by taking into account interference from all other routes.  For the interference-harnessing routing, on the other hand, we select a relay (at stage $k$) to maximize the MIMO capacity defined by two consecutive stages $k-1$ and $k$. We provide a description of this algorithm for the network with two sources using Fig.~\ref{FIG-EH}. Generalization to the case with more sources is straightforward. We use the notation $\Cc_{{\rm MIMO}}(\{\mbox{R}_{1},\mbox{R}_{2}\},\{\mbox{R}_{3},\mbox{R}_{4}\})$ to denote the MIMO capacity induced by two transmitters $\{\mbox{R}_{1},\mbox{R}_{2}\}$ and two receivers $\{\mbox{R}_{3},\mbox{R}_{4}\}$. Without loss of generality, we focus on the path 1 (i.e., $i=1$ in Fig.~\ref{FIG-EH}). The algorithm proceeds as follows: 1) Establish the route from  $\mbox{S}_{1}$ to $\mbox{D}$ so that the number of hops is minimized and each link-capacity along the route is maximized (namely, the received power is maximized). This process closely follows the step that establishes the initial route in \cite{Hui}; 2) For the fixed route from $\mbox{S}_{1}$ to $\mbox{D}$, establish a new route from  $\mbox{S}_{2}$ to $\mbox{D}$, as illustrated by the steps in Fig.~\ref{FIG-EH} and described below:
\begin{itemize}
\item Find the relays within a communication range from the $\mbox{S}_{2}$ that have not been already chosen for other routes. The communication range is a design parameter to be determined as a function of a transmit power where the transmit power should be larger than a threshold to satisfy network connectivity \cite{Gupta}. Let $\Tc_{1}=\{\mbox{R}_{1,1},\ldots,\mbox{R}_{1,|\Tc_{1}|}\}$ denote the set of such relays (i.e., ``candidate relays" of the first stage). For the $j$-th relay in $\Tc_{1}$, we can compute the MIMO capacity (i.e., routing metric) $\Cc_{{\rm MIMO}} (\{\mbox{S}_{1},\mbox{S}_{2}\},\{\mbox{R}_{1,1}^{1},\mbox{R}_{1,j}\})$ where $\mbox{R}_{1,1}^{1}$ denotes the first relay on the route from $\mbox{S}_{1}$ to $\mbox{D}$. With $\mbox{S}_{1}$, $\mbox{S}_{2}$, and $\mbox{R}_{1,1}^{1}$ fixed, we choose a relay $\mbox{R}_{1,j}$ to maximize the MIMO capacity:
    \begin{align*}
    j_{1}^{\star}=\mbox{argmax}_{j \in [1:|\Tc_{1}|]} \Cc_{{\rm MIMO}} (\{\mbox{S}_{1},\mbox{S}_{2}\},\{\mbox{R}_{1,1}^{1},\mbox{R}_{1,j}\}).
    \end{align*}
\item Let $\mbox{R}_{1,1}^{2}=\mbox{R}_{1,j_{1}^{\star}}$. Find the candidate relays of the second stage, denoted by $\Tc_{2}=\{\mbox{R}_{2,1},\ldots,$ $\mbox{R}_{2,|\Tc_{2}|}\}$. Then, choose a relay $\mbox{R}_{2,j}$ to maximize the routing metric:
    \begin{align*}
    j_{2}^{\star}=\mbox{argmax}_{j \in [1:|\Tc_{2}|]} \Cc_{{\rm MIMO}} (\{\mbox{R}_{1,1}^{1},\mbox{R}_{1,1}^2\},\{\mbox{R}_{1,2}^{1},\mbox{R}_{2,j}\}),
    \end{align*} where $\mbox{R}_{1,2}^{1}$ denotes the second relay on the route from $\mbox{S}_{1}$ to $\mbox{D}$.
\end{itemize} By setting $\mbox{R}_{1,2}^{2}=\mbox{R}_{2,j_{2}^{\star}}$, we can establish the route from $\mbox{S}_{2}$ to the $\mbox{D}$. Then, for the fixed route from $\mbox{S}_{2}$ to $\mbox{D}$, update the route from $\mbox{S}_{1}$ to $\mbox{D}$. Repeat this process until negligible improvements in the sum throughput can be made.

Notice that since the performance of the proposed scheme is independent of the inter-path interference as explained in Section~\ref{subsec:coding}, we first apply the above procedure to find a path 1 and then apply the procedure to find a path 2 using only the relays not included in the path 1.

\begin{figure*}[t]
\centerline{\includegraphics[width=12cm]{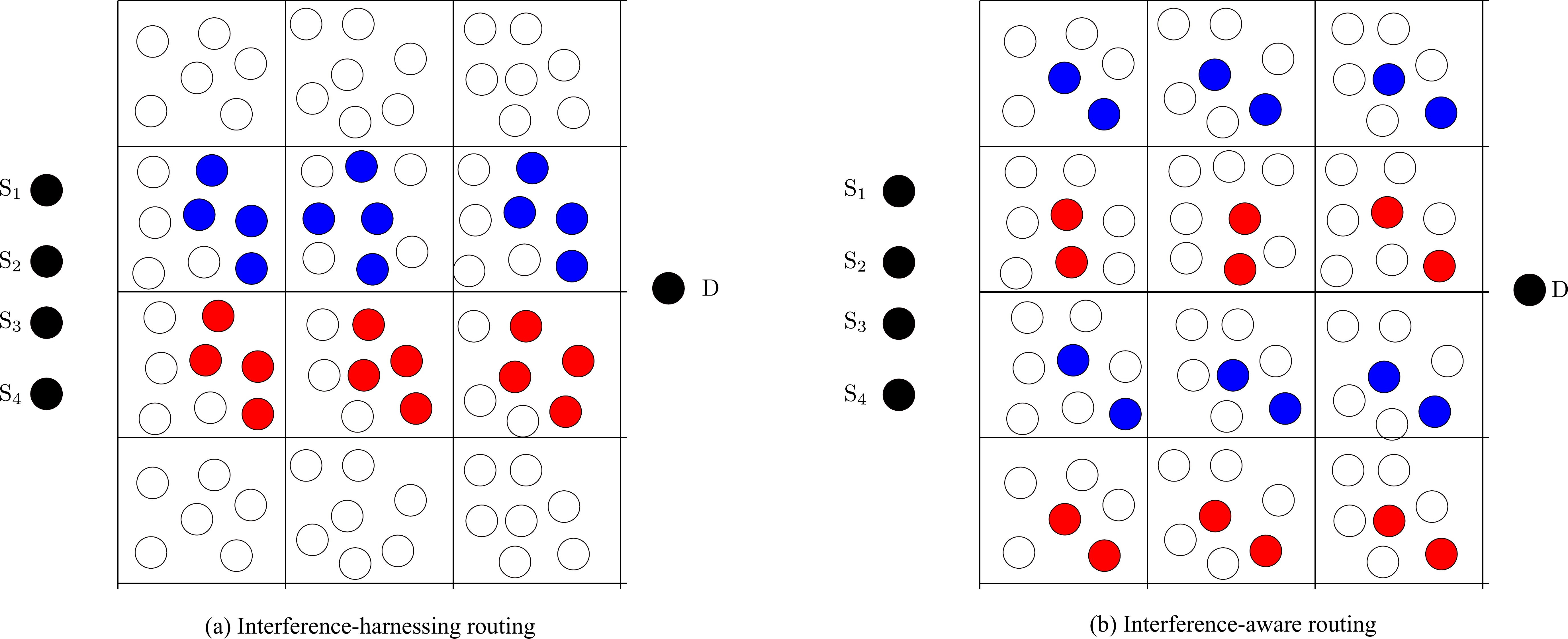}}
\caption{The opposite trend of the routes selected by interference-harnessing and interference-aware routing. Blue-colored relays establish the path 1 and red-colored relays establish the path 2, respectively.}
\label{FIG-EH-IA}
\end{figure*}

\begin{remark}\label{remark:EH} {\em (Energy harvesting)} In the example of Fig.~\ref{FIG-EH}, let $\Hm_{j}$ denote the $2 \times 2$ channel matrix between the two transmitters $\{\mbox{S}_{1},\mbox{S}_{2}\}$ and the two receivers $\{\mbox{R}_{1,1}^{1},\mbox{R}_{1,j}\}$ for some $\mbox{R}_{1,j} \in \Tc_{1}$, i.e.,
\begin{equation*}
\Hm_{j}=\left[
          \begin{array}{cc}
            h_{11} & h_{12} \\
            h_{j1} & h_{j2} \\
          \end{array}
        \right],
\end{equation*} where $h_{1i}$ denotes the channel from $\mbox{S}_{i}$ to $\mbox{R}_{1,1}^{1}$ along the established path 1 and  $h_{ji}$ denotes the channel from $\mbox{S}_{i}$ to a candidate relay $\mbox{R}_{1,j} \in \Tc_{1}$. Assuming that the transmit power is $\SNR$, we have:
\begin{align*}
&\Cc_{{\rm MIMO}} (\{\mbox{S}_{1},\mbox{S}_{2}\},\{\mbox{R}_{1,1}^{1},\mbox{R}_{1,j}\})= \log\det\left(\Id+\SNR\Hm_{j}\Hm_{j}^{\herm}\right)\\
&\;\;\;\stackrel{(a)}{\leq} \log(1+(|h_{11}|^2+|h_{12}|^2)\SNR)+ \log(1+(|h_{j1}|^2+|h_{j2}|^2)\SNR),
\end{align*} where (a) follows from the Hadamard's inequality and equality is achieved if the two columns of $\Hm_{j}$ are orthogonal. Thus, the obtained upper bound is maximized by choosing a relay $\mbox{R}_{1,j}$ to maximize $\mbox{SNRr}_{j}=(|h_{j1}|^2+|h_{j2}|^2)\SNR$. We observe that $\mbox{SNRr}_{j}$ is proportional to the power of the signal received from $\mbox{S}_{1}$ and $\mbox{S}_{2}$ at the $\mbox{R}_{1,j}$. We can use $\mbox{SNRr}_{j}$ as a routing metric approximating the MIMO capacity routing metric in the above iterative algorithm. This also implies that the interference-harnessing routing tends to choose a relay such that the signal power received from the previous stage is maximized.
\end{remark}

\begin{remark}{\em (Minimizing end-to-end delay):} For the interference-aware routing (based on DF), confining to nearest-neighbor transmissions maximizes the network throughput by mitigating the impact of inter-route interference \cite{Ozgur}. However, this approach increases the number of hops to reach destination thereby yielding long end-to-end delay. For the interference-harnessing routing, on the other hand, longer one-hop transmission (i.e., using a higher transmit power subject to a transmit power constraint) increases the network throughput by decreasing the number of relay stages $K$. This is due to the fact that, when relays use the optimized QMF, the throughput degrades as $K$ grows (see Figs.~\ref{SIM-SN} and~\ref{SIM-DN}). Thus, the interference-harnessing routing can be also more suitable for the systems with a delay constraint due to shorter end-to-end delay compared to the interference-aware routing.
\end{remark}

\begin{remark} {\em (Potential Gain of the interference-harnessing routing in multiple AgNs):}
A natural extension of our work is to consider wireless backhaul networks with multiple AgNs (say, $M$ AgNs). We can define a subnetwork $i$ consisting of AgN $i$ and the associated sources and relays for $i=1,\ldots,M$. Each subnetwork can be established via interference-harnessing routing. Then, the proposed transmission scheme can be applied to each subnetwork separately. In this case, there exist {\em inevitable} interference caused by the relays associated with different subnetworks. We refer to such interference as {\em inter-network} interference. Due to the use of the interference-harnessing routing, each subnetwork spans a narrow area over the entire network since in order to exploit interference, the routes tend to be chosen as closely as possible (see Fig.~\ref{FIG-EH-IA} (a)). For the interference-aware routing, on the other hand, each subnetwork spans a wide area over the whole network in order to avoid inter-route interference (see Fig.~\ref{FIG-EH-IA} (b)). Therefore, the interference-harnessing routing is much more efficient in avoiding the inter-network interference than the interference-aware routing. We expect that, when applying the proposed scheme to multiple AgNs, interference-harnessing routing further improves the performance compared with interference-aware routing. Moreover, we can employ the interference-harnessing routing to establish each subnetwork and the interference-aware routing to avoid inter-network interference. The algorithm establishes one subnetwork at a time while keeping other previously established subnetworks fixed, and repeats the process until negligible improvements in the sum throughput can be made. For the fixed subnetworks $i$ for $i \in \{1,\ldots,M\}\setminus \{j\}$, we can establish a subnetwork $j$ as follows:
\begin{itemize}
\item Perform the interference-aware routing to establish the first route of a subnetwork $j$, where each link-capacity on the route is computed by taking into account interference from all other subnetworks. This process can avoid inter-network interference.
\item Given the first route, perform the interference-harnessing routing to establish the subnetwork $j$.
\end{itemize}
\end{remark}

\begin{figure}[t]
\centerline{\includegraphics[width=10cm]{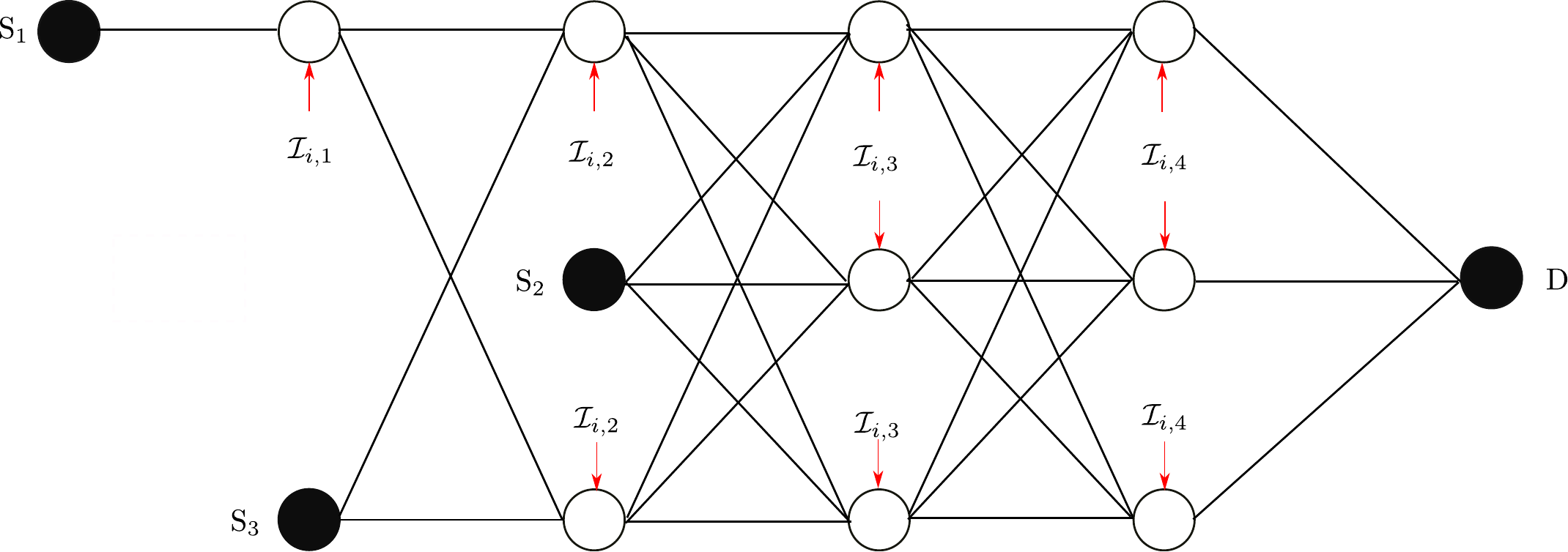}}
\caption{Equivalent simplified model of the asymmetric layered network with $K_{1}=4$, $K_{2}=2$, and $K_{3}=3$, where $K_{i}$ represents the number of relay stages so that the source $i$ can reach the destination.}
\label{FIG-ASYM}
\end{figure}

\begin{remark} {\em (asymmetric layered network):} One may be concerned that routes that roots from different sources will have different number of hops due to the various source-destination distances. We refer to such a network as an {\em asymmetric layered network}. This issue can be addressed by grouping the sources that are closely located and serving them simultaneously, which can be viewed as {\em user scheduling}. Furthermore, such user scheduling can maximize the ``interference-harnessing"\ gain since it is likely to produce a path such that the relays in each stage are closely located and hence each relay (using QMF) can collect more broadcasted energy. We emphasize that although
routes have different number of hops, the proposed scheme can be applied to such {\em asymmetric} layered networks as follows. Consider the example of Fig.~\ref{FIG-ASYM} where the source $i$ communicates with the destination with $(K_{i}+1)$ hops. For the asymmetric layered network, the destination performs the stage-by-stage successive decoding in Section~\ref{sec:scheme} in order to decode relays' messages as well as sources' messages. Until decoding the messages of relays at stage 3, the destination follows the same procedures as in Section~\ref{sec:scheme}. Then, using the $(\ell_{i,3}^{1},\ell_{i,3}^{2},\ell_{i,3}^{3})$ and the side-information $\Ic_{i,3}$, the destination can decode the two relays' messages
$(\ell_{i,2}^{1},\ell_{i,2}^{3})$ and the source 2's message $\underline{\wv}_{2}$. Similarly, using the $(\ell_{i,2}^{1},\ell_{i,2}^{3})$ and the side-information $\Ic_{i,2}$, the destination can decode the relay's message $\ell_{i,1}^{1}$ and the source 3's message $\underline{\wv}_{3}$. Finally, the destination can decode the source 1's message $\underline{\wv}_{1}$ using the $\ell_{i,1}^{1}$ and the side-information $\Ic_{i,1}$. This example shows that the proposed scheme can be naturally applied to the asymmetric layered networks.
\end{remark}


\section{Practical Construction of the Proposed Coding Scheme}\label{sec:decoding}

It was shown in Section~\ref{sec:Asymptotic} that the proposed scheme outperforms the MR. Recall that, in the proposed scheme, destination must perform joint decoding to decode relays' messages as well as sources' messages. It is noticeable that the joint decoding here is separately performed for each stage while it is done at once over entire messages in \cite{Avestimehr,Lim}. Nevertheless, the joint decoding typically requires higher complexity and makes it harder to design a practical coding scheme for the proposed scheme. Although some progresses of practical joint decoding have been made in \cite{Nagpal, Duarte} for small one relay network, we need further investigation to extend the previous works into the considered multihop network.

In this section, we develop a low-complexity decoder  named {\em stage-by-stage successive MIMO decoder} where a conventional MIMO decoding is applied to each stage instead of joint decoding. First of all, to avoid the joint decoder at the destination, we resort to the Wyner-Ziv quantization in which each relay $\mbox{R}_{i,k}^{j}$ chooses the quantization level such as
\begin{equation}\label{eq:defQ}
Q_{i,k}^{j} = \frac{1+\sum_{\ell=1}^{L}|h_{i,k}^{j,\ell}|^2\SNR}{2^{r_{i,k}^{j}}-1},
\end{equation} where $h_{i,k}^{j,\ell}$ denotes the $(j,\ell)$-th element of $\Hm_{i,k}$ and $r_{i,k}^{j}$ denotes the transmission rate of relay $\mbox{R}_{i,k}^{j}$. As noticed in Remark~\ref{remark:CF}, this particular choice of quantization level ensures that the destination can find a unique quantization sequence from the decoded relay's message $\ell_{i,k}^{j}$ and side-information.

Using the above fact, the destination performs the stage-by-stage successive MIMO decoding, i.e., the following procedures are repeated for $k=K+1,K,\ldots,1$, in that order:

\textbf{MIMO decoding at the stage $k$:} The destination has the side-information $\Ic_{i,k}$ and $\{\ell_{i,k}^{j}\}_{j=1}^{L}$ (i.e., decoded relays' messages at stage $k+1$).
\begin{itemize}
\item Using the side-information $\{\ell_{i,k}^{j}\}_{j=1}^{L}$ and $\Ic_{i,k}$, the destination can find {\em unique} quantized sequences $\{\hat{\underline{\yv}}_{i,k}^{j}\}_{j=1}^{L}$:
\begin{align*}
\hat{\underline{\yv}}_{i,k}^{j} &= \sum_{\ell=1}^{L} h_{i,k-1}^{j,\ell}\underline{\xv}_{i,k-1}^{\ell}+ \underbrace{\sum_{\ell=1}^{L} g_{i,k}^{j,\ell} \underline{\xv}_{i,k+1}^{\ell} + \sum_{\ell=1}^{L} s_{i,k}^{j,\ell} \underline{\xv}_{\bar{i},k}^{\ell}}_{\Ic_{i,k}}+ \hat{\underline{\zv}}_{i,k}^{j} + \underline{\zv}_{i,k}^{j},
\end{align*} and where $\hat{\underline{\zv}}_{i,k}^{j}$ denotes an i.i.d. Gaussian random variable $\sim \Cc\Nc(0,Q_{i,k}^{j})$.
\item After removing the known interference $\Ic_{i,k}$, the destination can decode relays' messages $(\ell_{i,k-1}^{1},\ldots,\ell_{i,k-1}^{L})$ from the resulting ``quantized MIMO MAC" channel, given by
\begin{align*}
\left[
  \begin{array}{c}
    \tilde{\underline{\yv}}_{i,k}^{1} \\
    \vdots \\
    \tilde{\underline{\yv}}_{i,k}^{L} \\
  \end{array}
\right] &= \Hm_{i,k}\left[
  \begin{array}{c}
    \underline{\xv}_{i,k-1}^{1}(\ell_{i,k-1}^{1}) \\
    \vdots \\
    \underline{\xv}_{i,k-1}^{L}(\ell_{i,k-1}^{L}) \\
  \end{array}
\right] +\left[
  \begin{array}{c}
    \hat{\underline{\zv}}_{i,k}^{1}+\underline{\zv}_{i,k}^{1} \\
    \vdots \\
    \hat{\underline{\zv}}_{i,k}^{L}+\underline{\zv}_{i,k}^{L} \\
  \end{array}
\right],
\end{align*} where $\tilde{\underline{\yv}}_{i,k}^{j} = \hat{\underline{\yv}}_{i,k}^{j}-\Ic_{i,k}$. Equivalently, we have the matrix form as
\begin{align}
\tilde{\underline{\Ym}}_{i,k} = \Hm_{i,k} \underline{\Xm}_{i,k-1} + \hat{\underline{\Zm}}_{i,k} + \underline{\Zm}_{i,k},\label{eq:channelO}
\end{align} where each row of a matrix consists of $n$-dimensional vector.
\end{itemize}

Since each stage $k$ consists of the {\em quantized} MIMO MAC defined in (\ref{eq:channelO}), many approaches to MIMO decoding can be applied. Clearly, the best performance can be attained by maximum likelihood (ML) decoding, which actually achieves the information-theoretical rates of using Wyner-Ziv quantization in Figs.~\ref{SIM-SN} and~\ref{SIM-DN}. The complexity of this approach, however, is exponential in the product of the coding blocklength $n$ and the number of receiver antennas (i.e., $L$ in our case). The complexity of ML decoding can be significantly reduced through the use of sphere decoding algorithms \cite{Viterbo, Damen}. Rather than processing all the observed signals from the antennas jointly, one simple and widely-used approach is to separate out the transmitted data streams using linear equalization and then decode each data stream individually such as zero-forcing (ZF), linear minimum mean-squared error (MMSE), and integer-forcing (IF) receivers \cite{Zhan}. The ZF receiver (a.k.a., decorrelator) inverts the channel matrix so that each data stream can be recovered via a single-user decoder. The MMSE receiver performs the same operation except with a {\em regularized} channel inverse that accounts for possible noise amplification. Both of these architectures permit the use of powerful point-to-point channel codes (e.g., Turbo code, LDPC code, and Polar code) that can achieve high data rates at practically-relevant SNRs. Recently, IF receiver was presented in \cite{Zhan} where the decoder first eliminates the noise by decoding a linear combination of interfering data streams and then eliminates interference between data streams in the digital domain. This scheme is based on the fact that each data stream is drawn from the same lattice codebook which ensures that any integer combination of codewords is itself a codeword, and thus decodable at high rates. Further, low-complexity coding frameworks based on QAM modulation and non-binary linear codes haven been proposed in \cite{Feng,Hong}.

In the next, we attain the performance of the low-complexity scheme with linear receivers. In this case, the receiver (i.e., destination)
applies the equalization matrix $\Bm_{i,k} \in \Cc^{L \times L}$ to obtain
\begin{equation}
\tilde{\underline{\Ym}}_{i,k}' = \Bm_{i,k}\underline{\Hm}_{i,k}\underline{\Xm}_{i,k} + \Bm(\hat{\underline{\Zm}}_{i,k} + \underline{\Zm}_{i,k}),
\end{equation} where $\Bm_{i,k}$ will be specifically defined according to linear schemes. From this, we can derive an achievable rate $r_{i,k}^{j}$ as a function of $\Bm_{i,k}$:
\begin{align*}
r_{i,k}^{j} =\log\left(1+\frac{\SNR\left|\underline{\bv}_{i,k}^{j}\hv_{i,k}^{j}\right|^2}{\sum_{\ell=1}^{L}|b_{i,k}^{j,\ell}|^2(1+Q_{i,k}^{\ell})+\SNR\sum_{\ell \neq j}\left|\underline{\bv}_{i,k}^{j}\hv_{i,k}^{\ell}\right|^2}\right),
\end{align*} where $\hv_{i,k}^j$ denotes the $j$-th column of $\Hm_{i,k}$ and $\underline{\bv}_{i,k}^{j}=(b_{i,k}^{j,1},\ldots,b_{i,k}^{j,L})$ denotes the $j$-th row of $\Bm_{i,k}$. With initial value $Q_{i,K+1}^{\ell}=0$ for all $i$ and $\ell$, we can derive all relays' rates at stage $k$ for $k=K,\ldots,1$ and sources' rates recursively. From \cite{Zhan}, we classify the equalization matrix $\Bm_{i,k}$ according to linear schemes: i) $\Bm_{i,k} = \Hm_{i,k}^{-1}$ for ZR receiver; ii) $\Bm_{i,k} = \Hm_{i,k}^{\herm}(\Qm_{i,k}^{-1} + \Hm_{i,k}\Hm_{i,k}^{\herm})^{-1}$ for linear MMSE receiver, where $\Qm_{i,k}$ denotes a diagonal matrix with $\SNR/(1+Q_{i,k}^{j})$ as its $j$-th entry; iii) $\Bm_{i,k} = \Am_{i,k}\Hm_{i,k}^{\herm}(\Qm_{i,k}^{-1} + \Hm_{i,k}\Hm_{i,k}^{\herm})^{-1}$ for IF receiver, where $\Am_{i,k} \in \ZZ[j]^{L \times L}$ is an integer full-rank matrix and is optimized as a function of channel matrix, and $\Qm_{i,k}$ denotes a diagonal matrix with its $j$-th diagonal element $\SNR/(1+Q_{i,k}^{j})$.
%
%
If we choose $\Am=\Id$, this scheme reduces to linear MMSE receiver. The key step underlying this approach is the selection of an integer matrix $\Am$ to approximate the channel matrix $\Hm_{i,k}$. Although finding the optimal $\Am_{i,k}$ has a worst-case complexity that is exponential in $L$, this search only needs to be performed once per coherence interval. In practice, efficient approximation algorithms (see \cite{LLL} and \cite{Hong} for details) can be used to find near-optimal $\Am_{i,k}$ in polynomial time.

\section{Numerical Results: Non-Asymptotic Case}\label{sec:SIM}


\begin{figure}[t]
\centerline{\includegraphics[width=10cm]{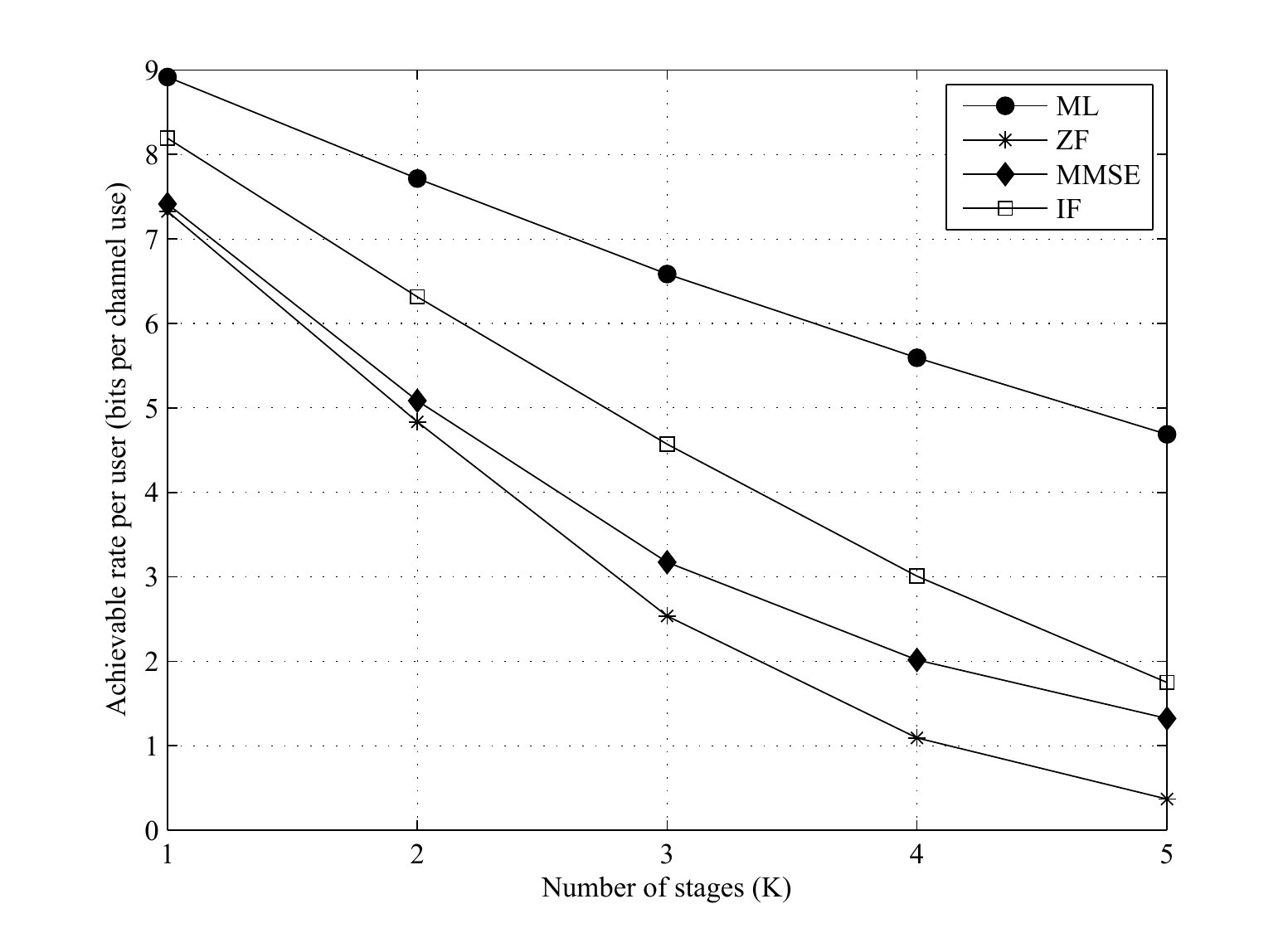}}
\caption{$\SNR=30$dB and $L=4$. Performance comparison of the proposed low-complexity scheme for various receiver architectures as a function of the number of stages $K$.}
\label{SIM-LR}
\end{figure}

\begin{figure}[t]
\centerline{\includegraphics[width=10cm]{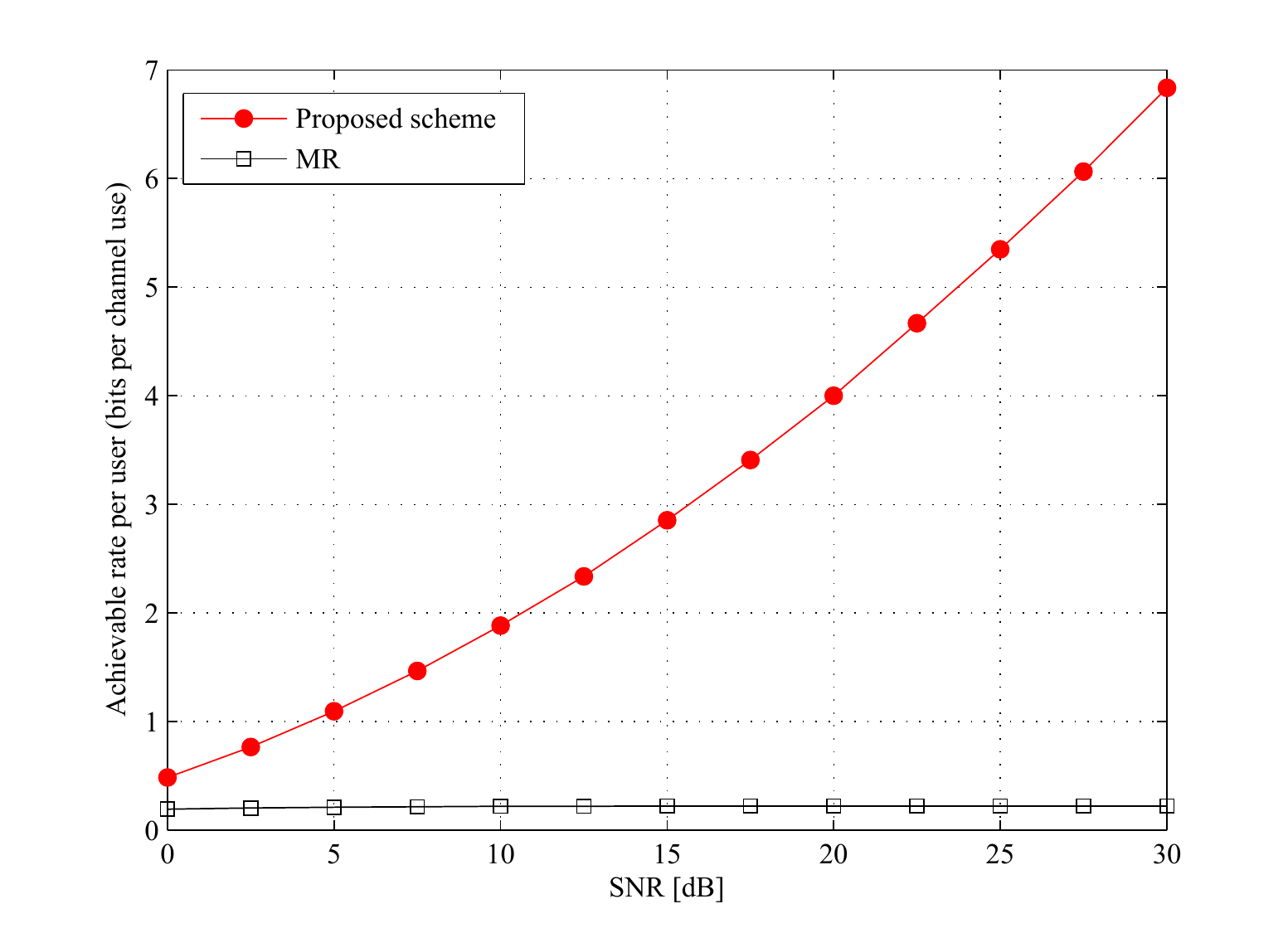}}
\caption{$L=4$ and $K=3$. Performance comparisons of the proposed scheme and multihop routing where the low-complexity decoding (with ML) is used for the proposed scheme.}
\label{SIM-SNR}
\end{figure}

\begin{figure}[t]
\centerline{\includegraphics[width=10cm]{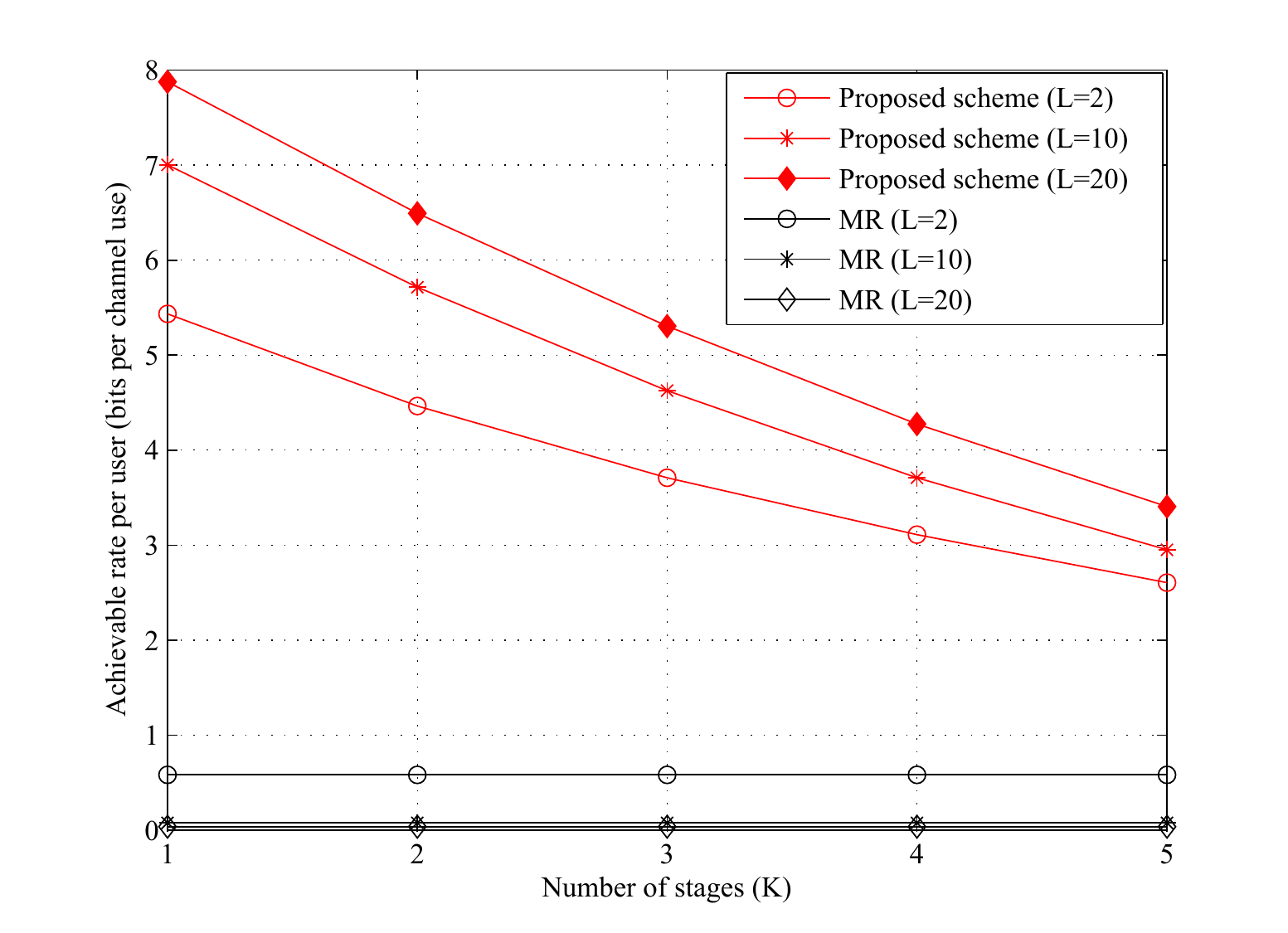}}
\caption{$\SNR=20$dB. Performance comparisons of the proposed scheme and multihop routing where the low-complexity decoding (with ML) is used for the proposed scheme.}
\label{SIM-HOPS}
\end{figure}

In this section, we evaluate the performance of the proposed low-complexity scheme in Section~\ref{sec:decoding} for non-asymptotic cases (i.e., small number of users). From this, we can identify the practical feasibility of the proposed scheme. We first investigate the performance of the proposed scheme for various receiver architectures namely ML, ZF, MMSE, and IF receivers. In our simulation, we consider the dense network with i.i.d. channel matrix $\Hm_{i,k}$ where each entry has a complex Gaussian distribution with zero mean and unit variance. Notice that in Fig.~\ref{SIM-LR}, ML receiver achieves the information-theoretical achievable rate. Among the linear receivers, IF receiver shows the best performance achieving about 1-bit gain over MMSE receiver. For $K < 3$, IF receiver can provide a satisfactory performance within 1-bit from the ML performance, significantly reducing the decoding complexity over ML receiver. Thus, IF receive can be a good candidate for a practical system where the number of stages is limited due to a delay constraint. For $K \geq 4$, however, a low-complexity ML receiver (e.g., sphere decoding) should be considered.

We next evaluate the performance of interference-harnessing algorithm and compare it to the performance of interference-aware routing. In our simulation, we consider the multihop wireless network depicted in Fig.~\ref{FIG-EH-IA} with $L=4$ sources and $K=3$ stages of relays. We define the {\em cluster} (i.e., a square in Fig.~\ref{FIG-EH-IA}) whose area depends on the transmit power. It is assumed that each relay can only communicate with the relays within eight neighboring clusters and its own cluster (i.e., communication range). Any interference outside the communication range is captured by additive Gaussian noise. We further assume the symmetric channel model for which a channel coefficient (within the communication range) is defined as $\SNR\exp(j\theta)$ where $\SNR$ captures the distance-dependent path-loss and transmit power, and $\theta \sim \mbox{Unif}[0,2\pi)$ denotes a random i.i.d. phase. In our simulation, we assume that the whole network is divided into $4 \times K$ disjoint clusters as shown in Fig.~\ref{FIG-EH-IA} with $K=3$, where $K$ captures the source-destination distance (i.e., determine the network area). It is also assumed that each cluster contains the $n_{c}$ half-duplex relays with $n_{c} \geq L$. Notice that for a given $K$ (i.e., network area), the network is denser and denser as $L$ grows, since the network includes more nodes. Hence, the parameters $K$ and $L$ control the network area and the network density, respectively.

For such network, the interference-harnessing and interference-aware routing are used to establishes the $L$ routes (per each path) for the proposed scheme and the MR, respectively. That is, the iterative algorithm in \cite{Hui} is applied to both cases with different routing metrics. For the interference-harnessing routing, we employ the simple routing metric in Remark~\ref{remark:EH} in which we only need to compute the received power at each candidate relay and choose a relay with the maximum received power. In the example of Fig.~\ref{FIG-EH-IA}, the relays chosen by the algorithm are shown in the blue and red circles. This can be easily extended into a general $L$ due to the assumption of $n_{c} \geq L$. To be specific, the interference-harnessing routing chooses the 6 clusters as in Fig.~\ref{FIG-EH-IA} (a) where each active cluster contains the $L$ selected relays while the interference-aware routing chooses the 12 clusters as in Fig.~\ref{FIG-EH-IA} (b) where each active cluster contains the $L/2$ selected relays. For the MR, there are two types of relays according to the number of received interference signals where the relays in the top or bottom route receive the $(\frac{3}{2}L - 2)$ interference signals and the relays in the middle routes receive the $(2L-2)$ interference signals. Therefore, the achievable symmetric rate of the MR is determined by the minimum of all message rates, given by
\begin{equation*}
r_{{\rm MR}} = \log\left(1+\SNR/(1+(2L-2)\SNR)\right).
\end{equation*} Fig.~\ref{SIM-SNR} shows that the proposed scheme outperforms the MR having a larger performance gain as $\SNR$ increases. This is because the strong interference limits the performance of the MR while it further improves the performance of the proposed scheme. From Fig.~\ref{SIM-HOPS}, we observe that in contrast to the MR, the performance of the proposed scheme is improved as the number of users $L$ increases, which is well-matched to the asymptotic result in Section~\ref{subsec:DN}.


\section{Conclusion}\label{sec:conclusion}

We presented a novel transmission scheme for wireless backhaul networks. In our scheme, $L$ sources transmit their messages over multiple hops using $L$ layers at each hop. In contrast to routing, in our scheme, message from every source is simultaneously forwarded by all $L$ relays at every hop. This is achieved by developing QMF/NNC by which a relay forwards a function of all source messages simultaneously.

We optimized both the relay selection as well as the QMF scheme developed by the relays. In particular, we proposed the optimized QMF that performs optimal quantization of received signals at the relays. As we showed, the optimized QMF significantly outperforms other forms of QMF as well as interference-aware routing (based on decode-and-forward) in both sparse and dense networks. Furthermore, the obtained performance gain increases as the network becomes denser (i.e., load becomes higher). An interesting result that came out of our analysis is that the energy efficiency of the proposed scheme increases as the network becomes denser. Based on this result, we proposed the algorithm to select the relays that we referred to as {\em interference-harnessing} routing. In interference-harnessing routing, in contrast to the interference-aware routing, the routing criterion selects relays to exploit interference, instead to avoid it, thereby greatly increasing the energy efficiency in the network. By using interference-harnessing routing and the proposed QMF, the rate performance of multihop routing can be achieved by significantly reducing transmit power at each relay.

Furthermore, to overcome the drawback of QMF/NNC which depends on prohibitively complex joint decoding at the destination, we developed a low-complexity stage-by-stage successive MIMO decoding. The proposed decoder can be implemented using a conventional MIMO decoding as ML, ZF, linear MMSE, and IF receivers. By comparing these various receiver architectures, we identified that, when the number of hops is less than 4, IF receiver provides a satisfactory performance within a 1-bit from the ML performance and shows much better performance than multihop routing. In practice, the maximum number of hops is usually limited due to a delay constraint. For such case, the proposed scheme can provide a substantial gain over the multihop routing with similar encoding/decoding complexity, by implementing our scheme with IF linear receiver and a point-to-point channel code (e.g., Turbo code, LDPC code, and Polar code).



\begin{thebibliography}{1}

\bibitem{Pi} Z. Pi and F. Khan, ``An introduction to millimeter-wave mobile broadband systems," {\em IEEE Commun. Mag.,} vol. 49, pp. 101-107, Jun. 2011.


\bibitem{SmallCell} Small Cell Forum, ``Backhaul technologies for small cells," white paper, document 049.02.01, Dec. 2013.


\bibitem{Hui} D. Hui and J. Axnas, ``Joint Routing and Resource Allocation for Wireless Self-Backhaul in an Indoor Ultra-Dense Network," in {\em Proc. IEEE Int. Symp. Personal, Indoor and Mobile Radio Commun.}, pp. 3083-3088, London, UK, 2013.

\bibitem{Interdigital} ''Millimeter Wave Backhaul Mesh for Dense Small Cell Deployments," {\em Interdigital White Paper}, Jan. 2015.

\bibitem{Baldemair} R. Baldemair, T. Irnich, K. Balachandran, E. Dahlman, G. Mildh, Y. Sel\'en, S. Parkvall, M. Meyer, and A. Osseiran, ``Ultra-Dense Networks in Millimeter-Wave Frequencies," {\em IEEE Commun. Mag.,} vol. 53, pp. 202-208,  Jan. 2015.


\bibitem{Daniels} R. Daniels, R. Heath, J. Murdock, and T. Rappaport, {\em 60 GHz Wireless Communication Systems} Prentice Hall Press, 2012.


\bibitem{Jain} K. Jain, J. Padhye, V. Padmanabhan, and L. Qiu, ``The Impact of Interference on Multi-Hop Wireless Network Performance," in {\em Proc. The Annual Int. Conf.  Mobile Computing and Networking (MobiCom),} pp. 66-80, Miami, Florida, Sep. 2013.


\bibitem{Draves} R. Draves, J. Padhye, and B. Zill, ``Routing in Multi-Radio, Multi-Hop Wireless Mesh Networks," in {\em Proc. The Annual Int. Conf. Mobile Computing and Networking (MobiCom),} pp. 114-128, Maui, Hawaii, Sep. 2014.


\bibitem{Parissidis} G. Parissidis, M. Karaliopoulos, T. Spyropoulos, and B. Plattner, ``Interference-Aware Routing in Wireless Multihop Networks," {\em IEEE Trans. Mobile Computing,} vol. 10, pp. 716-733, May 2011.


\bibitem{Park} S.-H. Park, O. Simeone, O. Sahin, and S. Shamai, ``Multihop Backhaul Compression for the Uplink of Cloud Radio Access Networks," in {\em Proc. IEEE Int. Symp. Inf. Theory (ISIT),} Honolulu, HI, Jun-Jul. 2014.
%


\bibitem{Avestimehr} S. Avestimehr, S. Diggavi, and D. Tse, ``Wireless network information flow: A deterministic approach," {\em IEEE Trans. Inf. Theory,} vol. 57, pp. 1872-1905, Apr. 2011.
\bibitem{Lim} S. Lim, Y. H. Kim, A. E. Gamal, and S. Chung, ``Noisy Network Coding," {\em IEEE Trans. Inf. Theory,} vol. 57, pp. 3132-3152.

\bibitem{Hou} J. Hou and G. Kramer, ``Short MEssage Noisy Network Coding with a Decode-Forward Option," {\em submitted to IEEE Trans. Inf. Theory,} Aug. 2013.


\bibitem{Peters} S. W. Peters and R. W. Heath, Jr., ``The future of WiMAX: Multi-hop relaying with IEEE 802.16j," {\em IEEE Commun. Mag.} vol. 1, pp. 104-111, Jan. 2009.

\bibitem{Peters1} S. W. Peters, A. Y. Panah, K. T. Truong, R. W. Heath, Jr., ``Relay Architectures for 3GPP LTE-Advanced," {\em EURASIP Journal on Advances in Signal Processing,} 2009.

\bibitem{Parkvall} S. Parkvall, E. Dahlman, A. Furuskar, Y. Jading, M. Olsson, S. Wanstedt, and K. Zangi, ``LTE-Advanced - Evolving LTE towards IMT-Advanced," in {\em Proc. IEEE Vech. Tech. Conf.,} Calgary, BC, Sept. 2008.


\bibitem{Peyman} P. Razaghi, S.-N. Hong, L. Zhou, W. Yu, and G. Caire, ``Two Birds and One Stone: Gaussian Interference Channel With a Shared Out-of-Band Relay of Limited Rate," {\em IEEE Trans. Inf. Theory,} vol. 59, no. 7, pp. 4192-4212, Jul. 2013.



\bibitem{Sanderovich} A. Sanderovich, O. Somekh, H. V. Poor, and S. Shamai (Shitz), ``Uplink Macro Diversity of Limited Backhaul Cellular Network," {\em IEEE Trans. Inf. Theory,} vol. 55, pp. 3457-3478, Aug. 2009.


\bibitem{Kolte} R. Kolte, A. Ozgur, and A. E. Gamal, ``Capacity Approximations for Gaussian Relay Networks," [Online] http://arxiv.org/abs/1407.3841.


\bibitem{Cover} T. Cover and J. Thomas, {\em Elements of Information Theory}. John Wiley Sons, Inc., 1991.


\bibitem{Gupta} P. Gupta and P. R. Kumar, ``Critical power for asymptotic connectivity," in {\em Proc. IEEE Conf. Decision and Control,} Tampa, FL, Dec. 1998.



\bibitem{Wyner} A. D. Wyner, ``Shannon-theoretic approach to a Gaussian cellular multiple-access channel," {\em IEEE Trans. Inf. Theory,} vol. 40, pp. 1713-1727, Nov. 1994.



\bibitem{Hong-SC} S.-N. Hong and G. Caire, ``Beyond Scaling Law: On the Rate Performance of Dense Device-to-Device Wireless Networks,"  {\em to appear in IEEE Trans. Inf. Theory,} 2015.



\bibitem{Ozgur} A. Ozgur, O. Leveque, and D. Tse, ``Operating Regimes of Large Wireless Networks," Foundations and Trends in Networking, 2011.


\bibitem{Nagpal} V. Nagpal, I.-H. Wang, M. Jorgovanovic, D. Tse, and B. Nikloi'c, ``Coding and System Design for Quantize-Map-and-Forward Relaying," {\em IEEE Jour. Sel. Areas in Commun.,} vol. 31, pp. 1423-1435, Aug. 2013.
\bibitem{Duarte} M. Duarte, A. Sengupta, S. Brahma, C. Fragouli, and S. Daggavi, ``Quantize-map-forward (QMF): an experimental study," in {\em Proc. the fourteenth ACM int. symp.  Mobile ad hoc networking and computing (MobiHoc),} pp. 227-236, New York, NY, 2013.

\bibitem{Viterbo} E. Viterbo and J. Boutros, ``A universal lattice decoder for fading channels," {\em IEEE Trans.  Inf. Theory,} vol. 45, pp. 1639-1642, Jul. 1999.
\bibitem{Damen} M. O. Damen, H. E. Gamal, and G. Caire, ``On maximum-likelihood detection and the searc hfor the closest lattice point," {\em IEEE Trans. Inf. Theory,} vol. 49, pp. 2389-2402, Oct. 2003.


\bibitem{Zhan} J. Zhan, B. Nazer, U. Erez, and M. Gastpar, ``Integer-Forcing Linear Receivers," {\em IEEE Trans. Inf. Theory}, vol. 60, pp. 7661-7685, Dec. 2014.
\bibitem{Feng} C. Feng, D. Silva, and F. Kschischang, ``An algebraic approach to physical-layer network coding," {\em IEEE Trans. Inf. Theory,} vol. 59, pp. 7576-7596, Nov. 2013.



\bibitem{LLL} A. K. Lenstra, H. W. Lenstra, and L. Lov'saz, ``Factoring polynomials with rational coefficients," {\em Mathematische Annalen}, vol. 261, pp. 515-534, 1982.

\bibitem{Hong} S.-N. Hong and G. Caire, ``Compute-and-Forward Strategies for Cooperative Distributed Antenna Systems,” {\em IEEE
Trans. Inf. Theory,} vol. 59, pp. 5227-5243, Aug. 2013.


\end{thebibliography}
\end{document}